\documentclass[12pt]{article}
\usepackage{amsthm, bm, graphicx, hyperref, mathrsfs, bbm}
\usepackage{amsfonts}
\usepackage{amsmath}
\usepackage{amssymb}
\usepackage{dsfont}
\usepackage{graphicx}
\usepackage{color}
\usepackage{braket}
\usepackage{epstopdf}
\usepackage[footnotesize]{caption}
\usepackage{amsthm}
\usepackage{enumitem}
\usepackage{mathrsfs}
\usepackage{scalerel}
\usepackage{mathtools}     
\usepackage{subeqnarray}         
\usepackage{cases}               
\usepackage{color}
\usepackage{subfigure}
\usepackage{cite}               
\usepackage{hyperref}            
\usepackage{multirow,makecell}   
\usepackage{textcomp}
\usepackage{wasysym}
\usepackage{url}
\usepackage[margin=3cm]{geometry}
\usepackage{geometry}
\usepackage{amsfonts}
\usepackage{graphicx}
\usepackage{float}
\usepackage{amsmath}
\usepackage{hyperref}
\usepackage{tensor}
\usepackage{subfigure}
\usepackage{cases}
\usepackage{appendix}
\usepackage{multirow}
\usepackage{color}
\allowdisplaybreaks[3]

\newcommand*{\dif}{\mathop{}\!\mathrm{d}}

\begin{document}
\begin{titlepage}
\vspace{0.5cm}

\begin{center}
{\Large\bf{Holographic stress tensor correlators on higher genus Riemann surfaces}}
\lineskip .75em
\vskip 2.5cm
{\large{Song He$^{\clubsuit,\maltese,}$\footnote{hesong@jlu.edu.cn}, Yun-Ze Li$^{\clubsuit,}$\footnote{lyz21@mails.jlu.edu.cn}, Yunfei Xie$^{\clubsuit,}$\footnote{jieyf22@mails.jlu.edu.cn}}}
\vskip 2.5em
{\normalsize\it $^\clubsuit$Center for Theoretical Physics and College of Physics, Jilin University,\\
 Changchun 130012, People's Republic of China
\\$^{\maltese}$Max Planck Institute for Gravitational Physics (Albert Einstein Institute),\\ 
Am M\"uhlenberg 1, 14476 Golm, Germany}
\end{center}

\begin{abstract}
In this work, we present a comprehensive study of holographic stress tensor correlators on general Riemann surfaces, extending beyond the previously well-studied torus cases to explore higher genus conformal field theories (CFTs) within the framework of the Anti-de Sitter/conformal field theory (AdS/CFT) correspondence. We develop a methodological approach to compute holographic stress tensor correlators, employing the Schottky uniformization technique to address the handlebody solutions for higher genus Riemann surfaces. Through rigorous calculations, we derive four-point stress tensor correlators, alongside recurrence relations for higher-point correlators, within the $\mathrm{AdS}_3/\mathrm{CFT}_2$ context. Additionally, our research delves into the holography of cutoff $\mathrm{AdS}_3$ spaces, offering novel insights into the lower-point correlators of the $T\bar{T}$-deformed theories on higher genus Riemann surfaces up to the first deformation order.

\end{abstract}
\end{titlepage}

\newpage
\tableofcontents

\section{Introduction}
The Anti-de Sitter/conformal field theory (AdS/CFT) correspondence \cite{Maldacena:1997re,Gubser:1998bc,Witten:1998qj}, as a strong-weak duality, provides a powerful toolkit for understanding the behavior of strongly coupled quantum field theories. Especially, it offers us a way to compute the correlators of local operators in the boundary CFT by performing gravitational perturbative calculations in the bulk.\par
The correlators of local operators are the most fundamental observables of a CFT. Among them, the stress tensor correlators have received substantial attention. They contain information about the energy, momentum, and stress distribution of a system, enabling analyses of phenomena such as the c-theorem \cite{Zamolodchikov:1986gt} and others. Extensive research has been conducted on these correlators both within field theory and in the context of holography \cite{Liu:1998ty,DHoker:1999bve,Arutyunov:1999nw,Raju:2012zs,Friedan:1986ua,Eguchi:1986sb,He:2022jyt,Nguyen:2021pdz}. While CFTs on Riemann surfaces have been explored in depth, research on holographic correlators of the stress tensor has predominantly focused on CFTs with trivial topology. Further research into holographic field theories on manifolds with nontrivial topologies is necessary to provide nontrivial tests of $\text{AdS}_3/\text{CFT}_2$.\par
{In our previous works \cite{He:2023hoj,He:2024fdm},} we computed holographic torus correlators of the stress tensor by solving the boundary value problem of Einstein's equation in the bulk. The prescription we proposed applies to any Riemann surface. In \cite{He:2023knl} we applied our approach to compute holographic torus correlators involving both the scalar operator and the stress tensor. Additionally, we extended the procedure to holographic correlators at a finite cutoff in the bulk. We further extended our analysis to $\text{AdS}_5/\text{CFT}_4$ in \cite{He:2023wcs}, where we computed the holographic Euclidean thermal two-point correlators of the stress tensor and $\mathrm{U}(1)$ current from the $\mathrm{AdS}$ planar black hole. In this paper we will take another step beyond the torus: we consider the higher genus case.\par
The holography of arbitrary genus compact Riemann surfaces has long been established \cite{Krasnov:2000zq}. Moreover, higher genus partition functions have been previously investigated \cite{Yin:2007gv,Yin:2007at,Giombi:2008vd,Chen:2015uga}, both for the handlebody and non-handlebody solutions. Our paper focuses on the handlebody solution, which can be constructed through the Schottky uniformization, as outlined in \cite{Krasnov:2000zq}. Following the approach in \cite{He:2023hoj}, we calculate the holographic correlators of the stress tensor on the conformal boundary, based on the well-established near-boundary solution in the form of Fefferman-Graham coordinates \cite{fefferman1985conformal,Henningson:1998gx,deHaro:2000vlm,Skenderis:1999nb,fefferman2012ambient}. Our results coincide with the Ward identity of CFT on the general Riemann surface \cite{Eguchi:1986sb}, providing a non-trivial verification of $\mathrm{AdS}_3/\mathrm{CFT}_2$. We also derive recurrence relations for computing some higher-point correlators during the calculation.\par
Furthermore, we extend our procedure to the case of the cutoff-$\mathrm{AdS}_3/T\bar{T}\text{-CFT}_2$ correspondence. The $T\bar{T}$ deformation \cite{Zamolodchikov:2004ce,Smirnov:2016lqw}, as an integrable deformation, has attracted considerable attention in recent years. It has been proposed that the $\mathrm{AdS}_3$ gravity with a Dirichlet boundary as a cutoff at a finite radial coordinate is dual to a $T\bar{T}$-deformed $\mathrm{CFT}_2$ living on that Dirichlet boundary \cite{McGough:2016lol}. It is an interesting and valuable topic to calculate correlation functions of $T\bar{T}$-deformed theories. Stress tensor correlators of $T\bar{T}$-deformed CFTs have been investigated using various approaches \cite{Kraus:2018xrn,Hirano:2020nwq,He:2020udl,Li:2020pwa,Hirano:2020ppu,He:2022jyt,He:2023hoj,He:2023knl,He:2023kgq,He:2024fdm}. Nevertheless, most of relative studies focus on the $T\bar{T}$-deformed theory either on the complex plane or on the torus. The $T\bar{T}$ operator has been observed to lose factorization property in the presence of non-zero curvature \cite{Jiang:2019tcq}, posing a challenge for studying $T\bar{T}$-deformation in curved spacetime. However, the factorization property still holds in a large $c$ limit. Following the dynamical coordinate formulation in \cite{Caputa:2020lpa} established for $T\bar{T}$ deformation in curved spacetime with trivial topology, we generalize the construction to the case of general Riemann surfaces, which allows us to derive a set of flow equations describing how the modular parameters of the cutoff Riemann surface change along the flow. Based on this construction, we calculate the stress tensor one-point correlators and two-point correlators perturbatively, as a worthwhile attempt to study the $T\bar{T}$ deformation in curved spacetime with nontrivial topology.\par
The remainder of the paper is organized as follows. In section \ref{section Holography of Riemann surfaces}, we briefly review the Schottky uniformization as the basis for subsequent calculations. In section \ref{section Holographic correlators of higher genus CFT}, we calculate holographic correlators of the stress tensor on the conformal boundary. We review how to obtain holographic correlators through near-boundary analysis and the GKPW dictionary and calculate stress tensor one-point correlators in subsection \ref{subsection Holographic setup and one-point correlators}. Two-point correlators are computed in subsection \ref{subsection CFT two-point correlators}. Recurrence relations are derived in subsection \ref{subsection Recurrence relations and higher-point correlators}. In section \ref{section Holographic correlators at finite cutoff} we investigate holographic correlators at a finite cutoff. We derive the explicit form of the dynamical coordinate transformation for a Riemann surface cutoff in subsection \ref{subsection Dynamical coordinate transformation}. Perturbative stress tensor one-point and two-point correlators are calculated in subsection \ref{subsection Perturbative stress tensor one-point correlator} and \ref{subsection Perturbative stress tensor two-point correlator}, respectively. Section \ref{section Conclusions and perspectives} is for conclusions and perspectives. Additionally, we review the necessary definitions and properties of differentials and Green's function in appendix \ref{appendix Differentials and Green's function}, which are utilized in the main text. In the end, we list all the independent three-point and four-point correlators of the CFT case in appendix \ref{appendix List of three-point and four-point correlators}.

\section{Holography of Riemann surfaces}\label{section Holography of Riemann surfaces}
In this section, we briefly review the basics of the holography of arbitrary genus compact Riemann surfaces \cite{Krasnov:2000zq}. \par
In general, one can obtain all types of constant negative curvature three-dimensional spaces by identifying the points of the Euclidean AdS$_3$ appropriately. Specifically, starting with the Euclidean AdS$_3$, namely the three-dimensional hyperbolic space $H^3$, we can obtain other constant negative curvature space by the quotient construction
\begin{equation}
    H^3/\Sigma,
\end{equation}
where $\Sigma$ is the Kleinian group \cite{Krasnov:2000zq,maskit2012kleinian}, a discrete subgroup of $\mathrm{PSL}(2,\mathbb{C})$, the group of orientation-preserving isometries of $H^3$. The quotient presents some subtleties when extending the action of $\Sigma$ to the conformal boundary $S^2\simeq\mathbb{C}\cup\{\infty\}$: the action may have fixed points on $S^2$. We denote $\Lambda$ as the so-called limit set, that is, the closure of the set of fixed points of the action. The difference set $\Omega=S^2\backslash\Lambda$ is called the region of discontinuity of $\Sigma$, an open set on which the Kleinian group $\Sigma$ acts freely discontinuously. The quotient $\Omega/\Sigma$ then can be a smooth manifold, serving as the conformal boundary of the quotient space $H^3/\Sigma$.\par
The topology of the three-dimensional space obtained from quotient by a general Kleinian group can be very complicated \cite{maskit2012kleinian,Thurston+1997}. In this paper, we focus on the simplest case that $\Sigma$ is a classical Schottky group \cite{Krasnov:2000zq,Tuite:2019fqx,zograf1988uniformization}. In this case, the quotient space we obtain corresponds to a handlebody solution.\par
\begin{figure}[H]
    \centering
    \includegraphics[scale=0.30]{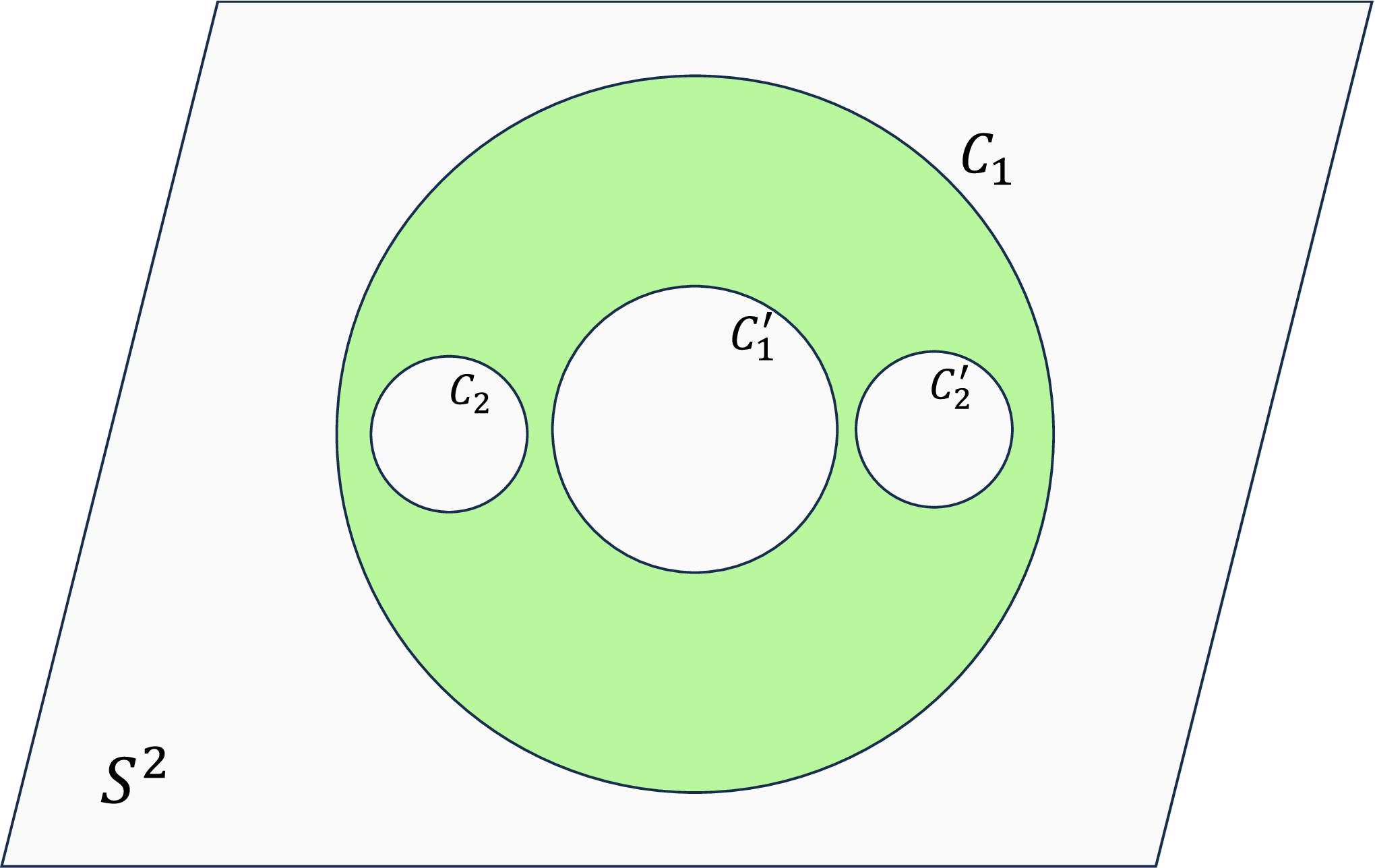}
    \caption{The illustration of the Schottky uniformization for a genus 2 Riemann surface. Schottky generator $L_i$ identifies the exterior of $C_i$ with the interior of $C_i'$, $i=1,2$, and vice versa. The green part stands for the fundamental domain $\mathcal{D}$ of the Schottky group $\Gamma_{g=2}$.}
    \label{figure 1}
\end{figure}

A marked classical Schottky group $\Gamma_g$ of genus $g$ is freely generated by $g$ loxodromic generators $L_1,\cdots,L_g$ (an element in $\mathrm{PSL}(2,\mathbb{C})$ is called loxodromic if its action has two fixed points on $S^2$ and $\mathrm{Tr}(L_i)\notin[0,2]$). To show how $\Gamma_g$ acts on the Riemann sphere, let $C_1,\cdots,C_g,C_1',\cdots,C_g'$ denote $2g$ non-intersecting circles in $S^2$. Each generator $L_i, i=1,2,\cdots,g$ can be represented in the form
\begin{equation}
    \frac{L_i(z)-a_i}{L_i(z)-b_i}=\lambda_i\frac{z-a_i}{z-b_i},\quad z\in\mathbb{C}\cup\{\infty\},
\end{equation}
where $a_i$ and $b_i$ are the two fixed points of $L_i$, which can always be chosen as the centers of $C_i$ and $C_i'$, and the multiplier $\lambda_i$ is a complex number with $0<|\lambda_i|<1$. The action of $L_i$ maps $C_i$ to $C_i'$, while the exterior of $C_i$ is mapped to the interior of $C_i'$, and vice versa. The fundamental domain of $\Gamma_g$ is thus the exterior of all the $2g$ circles. An example of the $g=2$ case is shown in figure \ref{figure 1}.\par
According to the classical retrosection theorem \cite{Koebe1914,hidalgo2008retrosection}, for any compact Riemann surface, one can always find a Schottky group $\Gamma$ such that the Riemann surface can be represented in the form $\Omega/\Gamma$, where $\Omega$ is the region of discontinuity of $\Gamma$. The Schottky group can be chosen such that the image of $C_i$'s become $g$ generators of the fundamental group $\pi_1(X)$ of the Riemann surface $X$ we construct. Also, the moduli space of $X$ can be obtained from $3g-3$ parameters of $\Gamma_g$ after fixing three parameters by Möbius transformation \cite{Krasnov:2000zq,zograf1988uniformization}. The procedure above is called Schottky uniformization.\par
The holography of Riemann surfaces has received significant attention in the study of 2+1 dimensional wormholes \cite{Aminneborg:1997pz,Brill:1998pr,Maldacena:2004rf,Skenderis:2009ju,Balasubramanian:2014hda}, where the handlebodies play a role of Euclidean counterparts of the Lorentzian wormholes in the sense of the real-time gauge/gravity duality \cite{Skenderis:2008dh,Skenderis:2008dg,Skenderis:2009ju}. Non-handlebody geometries are also considered in \cite{Maldacena:2004rf,Yin:2007at}.
\section{Holographic correlators of higher genus CFT}\label{section Holographic correlators of higher genus CFT}
In this section, we compute holographic correlators of the stress tensor of conformal field theories on Riemann surfaces of genus $g\geqslant2$. During the calculation, we follow the method developed in \cite{He:2023hoj,He:2023knl}. 
\subsection{Holographic setup and one-point correlators}\label{subsection Holographic setup and one-point correlators}
Firstly we review how to obtain holographic stress tensor correlators within the framework of $\mathrm{AdS}_3/\mathrm{CFT}_2$. As outlined in \cite{He:2023hoj}, we employ the Fefferman-Graham coordinates near the boundary\cite{fefferman1985conformal,Skenderis:1999nb} for the holographic calculation, in which the bulk metric can be expressed in the form
\begin{equation}
    \dif s^2=\frac{\dif \rho^2}{4\rho^2}+\frac{1}{\rho}g_{ij}(x,\rho)\dif x^i\dif x^j.\label{FGmetric}
\end{equation}
$g_{ij}(x,\rho)$ can be expanded into a series of $\rho$, which truncates in three dimension spacetime:
\begin{equation}
    g_{ij}(x,\rho)=g_{ij}^{(0)}(x)+g_{ij}^{(2)}(x)\rho+g_{ij}^{(4)}(x)\rho^2.\label{FGexpansion}
\end{equation}
With this metric, Einstein's equation can be reduced into three equations about $g^{(0)},g^{(2)}$ and $g^{(4)}$:
\begin{align}
    g^{(4)}_{ij}&=\frac{1}{4}g_{ik}^{(2)}g^{(0)kl}g^{(2)}_{lj},\label{EinEq1}\\
    \nabla^{(0)i}g_{ij}^{(2)}&=\nabla_j^{(0)}g^{(2)i}_{\quad i},\label{EinEq2}\\
    g^{(2)i}_{\quad i}&=-\frac{1}{2}R[g^{(0)}].\label{EinEq3}
\end{align}
In all three equations above, the covariant derivative and raising (lowering) indices are with respect to $g^{(0)}$, which can be identified with the boundary metric the CFT living in.\par
According to the GKPW dictionary of AdS/CFT\cite{Witten:1998qj}, there exists an equivalence between the bulk gravitational partition function and the generating functional of the boundary CFT, where the former can be approximated as a sum over all saddles in the semiclassical limit. In our calculation, we assume that only one saddle dominates, thus the generating functional of connected correlators of the CFT is equal to the on-shell action of this saddle. The one-point correlator of the stress tensor in the boundary CFT can be identified with the Brown-York tensor \cite{deHaro:2000vlm} in the case:
\begin{equation}
    \braket{T_{ij}}=-\frac{1}{8\pi G}(K_{ij}-Kh_{ij}+h_{ij}),
\end{equation}
where $G$ is Newton's constant{, which is related to the CFT central charge through the Brown-Henneaux relation \cite{Brown:1986nw} $c=\frac{3}{2G}$}. $K_{ij}$ and $h_{ij}$ are the extrinsic curvature and the induced metric of the boundary, respectively. Using the Fefferman-Graham coordinates and equations (\ref{EinEq1})(\ref{EinEq2})(\ref{EinEq3}) above, we can obtain the expression of one-point correlator with respect to $g^{(0)}$ and $g^{(2)}$
\begin{equation}
    \braket{T_{ij}}=\frac{1}{8\pi G}\left(g_{ij}^{(2)}-g^{(0)kl}g_{kl}^{(2)}g_{ij}^{(0)}\right)
\end{equation}
with the conservation law and holographic Weyl anomaly:
\begin{align}
    &\nabla^i\braket{T_{ij}}=0,\label{conservlaw}\\
    &\braket{T_i^i}=\frac{1}{16\pi G}R[g^{(0)}]\label{confanomaly}.
\end{align}\par
To compute multi-point correlators of the stress tensor, by definition, we only need to take the functional derivative of the one-point function with respect to the metric. Here we choose the convention of the definition to be
\begin{equation}
    \braket{T_{i_1j_1}(z_1)\cdots T_{i_nj_n}(z_n)}=-\frac{(-2)^n\delta^nI_{\mathrm{CFT}}}{\sqrt{\mathrm{det}(g^{(0)}(z_1))}\cdots\sqrt{\mathrm{det}(g^{(0)}(z_n))}\delta g^{(0)i_1j_1}(z_1)\cdots \delta g^{(0)i_nj_n}(z_n)}, \label{definition of stress tensor correlator}
\end{equation}
where $I_{\mathrm{CFT}}$ is the generating functional of connected correlators of the boundary CFT.\par
Now we return to the concrete calculation of higher genus correlators. we need to fix the boundary metric first. What should be noted is that in the higher genus case, the boundary metric cannot be set to be flat like the torus case. This can be seen from the Euler characteristic of a closed-oriented Riemann surface, given by $\chi=2-2g$, where $g\geqslant 2$ yields a negative value. Instead, a unique complete metric with constant negative curvature $R=-1$ \cite{zograf1988uniformization,Matone:1993tj} exists on any such compact Riemann surface. This metric can be obtained from the flat metric in conformal gauge via a Weyl transformation:
\begin{equation}
    \dif s^2=e^{2\phi(z,\Bar{z})}\dif z\dif\Bar{z},\label{boundarymetric}
\end{equation}
where $\phi(z,\Bar{z})$ is a Liouville field that satisfies
\begin{equation}
    8\partial_z\partial_{\Bar{z}}\phi=e^{2\phi}\label{Liouville equation}
\end{equation}
by the condition $R[g^{(0)}]=-1$. Moreover, to make the metric single-valued under the Schottky uniformization, the Liouville field $\phi$ must have the transformation property
\begin{equation}
    \phi(\gamma(z),\overline{\gamma(z)})=\phi(z,\bar{z})-\frac{1}{2}\ln|\gamma'(z)|^2\label{Lioutrans}
\end{equation}
under the action of any generator $\gamma$ of the Schottky group $\Gamma_g$. This property provides a boundary condition if we choose $\gamma$ as the generator $L_i$'s of $\Gamma_g$. The Liouville equation (\ref{Liouville equation}) is hard to solve with such quasiperiodic boundary conditions. There has been some work solving it numerically \cite{Maxfield:2016mwh, Wien:2017xzd}. Throughout this paper, we will keep the Liouville field $\phi$ in all expressions.\par
To obtain the correlators, we start by finding the proper Fefferman-Graham coordinates, in which the metric coincides with (\ref{boundarymetric}) on the boundary. Starting from the Poincaré coordinates
\begin{equation}
    \dif s^2=\frac{\dif\xi^2}{4\xi^2}+\frac{\dif y\dif\Bar{y}}{\xi},
\end{equation}
this can be done directly by taking the transformation\cite{Krasnov:2001cu}:
\begin{equation}
    \xi=\frac{\rho e^{-2\phi}}{(1+\rho e^{-2\phi}|\partial_z\phi|^2)^2},\quad\quad y=z+\partial_{\Bar{z}}\phi\frac{\rho e^{-2\phi}}{1+\rho e^{-2\phi}|\partial_w\phi|^2}.
\end{equation}
After the coordinate transformation, the metric becomes the Fefferman-Graham form {that} we need:
\begin{align}
    \dif s^2&=\frac{\dif \rho^2}{4\rho^2}+\frac{1}{\rho}e^{2\phi}\dif z\dif\Bar{z}+\mathcal{T}^\phi\dif z^2+\Bar{\mathcal{T}}^\phi\dif\Bar{z}^2+2\mathcal{R}\dif z\dif\Bar{z}\nonumber\\
    &\hspace{4cm}+\rho e^{-2\phi}(\mathcal{T}^\phi\dif z+\mathcal{R}\dif\Bar{z})(\Bar{\mathcal{T}}^\phi\dif\Bar{z}+\mathcal{R}\dif z),
\end{align}
where
\begin{equation}
    \mathcal{T}^\phi=\partial_z^2\phi-(\partial_z\phi)^2,\quad \mathcal{R}=\partial_z\partial_{\Bar{z}}\phi.
\end{equation}
Therefore, {when} compared with the general form of the metric in Fefferman-Graham coordinates (\ref{FGmetric})(\ref{FGexpansion}), we can read off
\begin{equation}
    \begin{split}
        &g^{(0)}=\begin{pmatrix}
        0&\frac{1}{2}e^{2\phi}\\\frac{1}{2}e^{2\phi}&0
        \end{pmatrix},\quad g^{(2)}=\begin{pmatrix}
        \mathcal{T}^\phi&\mathcal{R}\\\mathcal{R}&\Bar{\mathcal{T}}^\phi
        \end{pmatrix},\\
        &g^{(4)}=e^{-2\phi}\begin{pmatrix}
       \mathcal{T}^\phi \mathcal{R}&\frac{1}{2}(\mathcal{T}^\phi\Bar{\mathcal{T}}^\phi+\mathcal{R}^2)\\\frac{1}{2}(\mathcal{T}^\phi\Bar{\mathcal{T}}^\phi+\mathcal{R}^2)&\mathcal{R}\Bar{\mathcal{T}}^\phi
    \end{pmatrix}.
    \end{split}
\end{equation}
It is easy to verify that they indeed satisfy Einstein's equations (\ref{EinEq1})(\ref{EinEq2})(\ref{EinEq3}). Then we get the expression of one-point correlators in terms of the Liouville field:
\begin{equation}
    \begin{split}
        \braket{T_{zz}}&=\frac{1}{8\pi G}\mathcal{T}^\phi=\frac{1}{8\pi G}\left(\partial_z^2\phi-(\partial_z\phi)^2\right),\\
    \braket{T_{z\Bar{z}}}&=\braket{T_{\Bar{z}z}}=-\frac{\mathcal{R}}{8\pi G}=-\frac{1}{8\pi G}\partial_z\partial_{\Bar{z}}\phi=-\frac{e^{2\phi}}{64\pi G},\\
    \braket{T_{\Bar{z}\Bar{z}}}&=\frac{1}{8\pi G}\Bar{\mathcal{T}}^\phi=\frac{1}{8\pi G}\left(\partial_{\Bar{z}}^2\phi-(\partial_{\Bar{z}}\phi)^2\right).\label{1pt}
    \end{split}
\end{equation}
It's also straightforward to verify that they satisfy the conservation law (\ref{conservlaw}) and Weyl anomaly (\ref{confanomaly}).\par
There is also one point worth mentioning. Taking $\braket{T_{zz}}$ as an example, to ensure the holomorphic form $\braket{T_{zz}}\dif z^2$ is single-valued on the Riemann surface after Schottky uniformization, $\braket{T_{zz}}$ must be an automorphic form of type (2,0) \cite{Martinec:1986bq}, that is, $\braket{T_{zz}}$ must satisfy
\begin{equation}
    \braket{T_{zz}(\gamma(z))}\left[\gamma'(z)\right]^2=\braket{T_{zz}(z)}\label{automorphic verify1}
\end{equation}
on the covering space for any element $\gamma$ of the Schottky group. This can be verified straightforwardly. By utilizing the transformation property (\ref{Lioutrans}) of the Liouville field $\phi$, we can get
\begin{align}
    \braket{T_{zz}(\gamma(z))}=\frac{1}{8\pi G}\frac{1}{\left[\gamma'(z))\right]^2}\left(\partial_z^2\phi-(\partial_z\phi)^2-\frac{1}{2}S\{\gamma,z\}\right),\label{automorphic verify2}
\end{align}
where $S\{\gamma,z\}$ is the Schwarzian derivative
\begin{equation}
    S\{\gamma,z\}=\frac{\gamma'''(z)}{\gamma'(z)}-\frac{3}{2}\left(\frac{\gamma''(z)}{\gamma'(z)}\right)^2.\label{Schwarzian}
\end{equation}
This Schwarzian derivative term doesn't contribute because the Schottky group is a subgroup of $\mathrm{PSL}(2,\mathbb{C})$, and (\ref{Schwarzian}) vanishes for any $\gamma$ in $\mathrm{PSL}(2,\mathbb{C})$. Consequently, (\ref{automorphic verify2}) simplifies to the expected (\ref{automorphic verify1}).

\subsection{Two-point correlators}\label{subsection CFT two-point correlators}
Now we are ready to compute higher-order correlators of the stress tensor, starting with $\braket{T_{zz}(z)T_{ww}(w)}$ as an illustrative example. As previously mentioned, we initiate by varying the boundary metric:
\begin{equation}
    \delta g^{(0)}_{ij}\dif x^i\dif x^j=\epsilon\chi_{ij}\dif x^i\dif x^j.
\end{equation}
The variation of $g^{(0)}_{ij}$ induces a corresponding variation of the one-point correlator $\braket{T_{ij}}$, which can be formally expressed as a series of the infinitesimal parameter $\epsilon$:
\begin{equation}
    \sum\limits_{n=1}^\infty\epsilon^n\braket{T_{ij}}^{[n]}.
\end{equation}
We can solve $\braket{T_{ij}}^{[n]}$ order by order from the two equations (\ref{conservlaw}), (\ref{confanomaly}) and take $n$-th functional derivatives to obtain $(n+1)$-point correlators. For $\braket{T_{zz}(z)T_{ww}(w)}$, we take the first order {terms} of $\epsilon$ in (\ref{conservlaw}), and then take the functional derivative with respect to $\chi_{\Bar{z}\Bar{z}}$, and then evaluate the result in the unperturbed metric:
\begin{align}
    &\partial_{\Bar{z}}\frac{\delta\braket{T_{zz}}^{[1]}(z)}{\delta\chi_{\Bar{w}\Bar{w}}(w)}+\frac{1}{16\pi G}e^{-2\phi(z,\bar{z})}\big(12(\partial_z\phi)^2\partial_z+12\partial_z\phi\partial_z^2\phi-8(\partial_z\phi)^3-6\partial_z^2\phi\partial_z-6\partial_z\phi\partial_z^2\nonumber\\
    &\hspace{9cm}-2\partial_z^3\phi+\partial_z^3\big)\delta^{(2)}(z-w)=0.
\end{align}
This differential equation can be solved with Green's function on general Riemann surfaces\cite{Eguchi:1986sb,Martinec:1986bq,Tuite:2019fqx}. The Green's function is an automorphic form which satisfies
\begin{equation}
    \frac{1}{\pi}\partial_{\Bar{z}}G^z_{\ ww}(z,\Bar{z};w,\Bar{w})=\delta^{(2)}(z-w)-p_2(z,\bar{z};w),\label{differential equation for G}
\end{equation}
where
\begin{equation}
    p_2(z,\bar z;w)=\sum_{\alpha=1}^{3g-3}\mu_{\alpha\bar z}^z(z,\bar z)\phi_{\alpha ww}(w).
\end{equation}
Here, $\phi_{\alpha zz}$ and $\mu_{\alpha\Bar{z}}^z$ are holomorphic quadratic differential on the Riemann surface and Beltrami differential dual to $\phi_{\alpha zz}$\cite{Eguchi:1986sb, Tuite:2019fqx, DHoker:1988pdl}. We review the details about these differentials and Green's function on a general Riemann surface in appendix \ref{appendix Differentials and Green's function}. Thus we have
\begin{align}
    \frac{\delta\braket{T_{zz}(z)}}{\delta g^{(0)}_{\Bar{w}\Bar{w}}(w)}&=\frac{\delta\braket{T_{zz}}^{[1]}(z)}{\delta\chi_{\Bar{w}\Bar{w}}(w)}=\int_{\mathcal{D}}\dif^2z_0\,\delta^{(2)}(z_0-z)\frac{\delta\braket{T_{zz}}^{[1]}(z_0)}{\delta\chi_{\Bar{w}\Bar{w}}(w)}\nonumber\\
    &=\int_{\mathcal{D}}\dif^2z_0\,\left(\frac{1}{\pi}\partial_{\Bar{z}_0}G^{z_0}_{\ zz}(z_0,\Bar{z}_0;z,\Bar{z})+p_2(z_0,\bar{z}_0;z)\right)\frac{\delta\braket{T_{zz}}^{[1]}(z_0)}{\delta\chi_{\Bar{w}\Bar{w}}(w)},\label{two terms in TT}
\end{align}
where $\mathcal{D}$ is the fundamental domain of $\Gamma_g$. We consider the two terms separately. For the first term, integrating by part and applying Stokes' theorem, we obtain:
\begin{align}
    &\int_{\mathcal{D}}\dif^2z_0\,\frac{1}{\pi}\partial_{\Bar{z}_0}G^{z_0}_{\ zz}(z_0,\Bar{z}_0;z,\Bar{z})\frac{\delta\braket{T_{zz}}^{[1]}(z_0)}{\delta\chi_{\Bar{w}\Bar{w}}(w)}\nonumber\\
    =&\int_{\mathcal{D}}\dif^2z_0\,\frac{1}{\pi}\partial_{\Bar{z}_0}\left[G^{z_0}_{\ zz}(z_0,\Bar{z}_0;z,\Bar{z})\frac{\delta\braket{T_{zz}}^{[1]}(z_0)}{\delta\chi_{\Bar{w}\Bar{w}}(w)}\right]\nonumber\\
    &-\int_{\mathcal{D}}\dif^2z_0\,G^{z_0}_{\ zz}(z_0,\Bar{z}_0;z,\Bar{z})\Big[\frac{-1}{16\pi^2 G}e^{-2\phi(z_0,\bar{z}_0)}\big(12(\partial_{z_0}\phi)^2\partial_{z_0}+12\partial_{z_0}\phi\partial_{z_0}^2\phi-8(\partial_{z_0}\phi)^3\nonumber\\
    &-6\partial_{z_0}^2\phi\partial_{z_0}-6\partial_{z_0}\phi\partial_{z_0}^2-2\partial_{z_0}^3\phi+\partial_{z_0}^3\big)\delta^{(2)}(z_0-w)\Big]\nonumber\\
    =&\ -\frac{i}{2}\oint_{\partial \mathcal{D}}\dif z_0\,\frac{1}{\pi}G^{z_0}_{\ zz}(z_0,\Bar{z}_0;z,\Bar{z})\frac{\delta\braket{T_{zz}}^{[1]}(z_0)}{\delta\chi_{\Bar{w}\Bar{w}}(w)}-\frac{1}{16\pi^2 G}e^{-2\phi(w,\bar{w})}\partial_w^3G^w_{\ zz}(w,\Bar{w};z,\Bar{z}),
\end{align}
where $\partial\mathcal{D}$ denotes the boundary of the fundamental domain $\mathcal{D}$ of the Schottky group $\Gamma_g$. By the construction of Schottky uniformization, $\partial \mathcal{D}$ is just the $2g$ circles $C_i, C_i'$, $i=1,2,\cdots,g$. And $\forall z_i\in C_i$, there exists a $z_i'=\gamma_i(z_i)\in C_i'$. Remembering that both $G^{z_0}_{\ zz}(z_0,\Bar{z}_0;z,\Bar{z})$ and $\braket{T_{zz}}^{[1]}(z_0)$ are automorphic forms, we find
\begin{equation}
    \oint_{\partial D}=\sum\limits_{i=1}^g\left(\oint_{C_i}-\oint_{C_i'}\right)
\end{equation}
and
\begin{align}
    &\oint_{C_i'}\dif[\gamma_i(z^i_0)]\,G^{\gamma_i(z^i_0)}_{\ \ zz}(\gamma_i(z_0^i),\overline{\gamma(z_0^i)};z,\Bar{z})\frac{\delta\braket{T_{zz}}^{[1]}(\gamma_i(z^i_0))}{\delta\chi_{\Bar{w}\Bar{w}}(w)}\nonumber\\
    =&\ \oint_{C_i'}\dif z^i_0[\gamma'_i(z^i_0)]\,G^{z^i_0}_{\ \ zz}(z_0^i,\Bar{z}_0^i;z,\Bar{z})[\gamma'_i(z^i_0)]\frac{\delta\braket{T_{zz}}^{[1]}(z^i_0)}{\delta\chi_{\Bar{w}\Bar{w}}(w)}[\gamma'_i(z^i_0)]^{-2}\nonumber\\
    =&\ \oint_{C_i'}\dif z^i_0\,G^{z^i_0}_{\ \ zz}(z_0^i,\Bar{z}_0^i;z,\Bar{z})\frac{\delta\braket{T_{zz}}^{[1]}(z^i_0)}{\delta\chi_{\Bar{w}\Bar{w}}(w)}=\oint_{C_i}\dif z^i_0\,G^{z^i_0}_{\ \ zz}(z_0^i,\Bar{z}_0^i;z,\Bar{z})\frac{\delta\braket{T_{zz}}^{[1]}(z^i_0)}{\delta\chi_{\Bar{w}\Bar{w}}(w)}.
\end{align}
Therefore the boundary term cancels out, yielding:
\begin{align}
    &\int_{\mathcal{D}}\dif^2z_0\,\frac{1}{\pi}\partial_{\Bar{z}_0}G^{z_0}_{\ zz}(z_0,\Bar{z}_0;z,\Bar{z})\frac{\delta\braket{T_{zz}}^{[1]}(z_0)}{\delta\chi_{\Bar{w}\Bar{w}}(w)}=-\frac{1}{16\pi^2 G}e^{-2\phi(w,\bar{w})}\partial_w^3G^w_{\ zz}(w,\Bar{w};z,\Bar{z}).\label{TT2ptGreenpart}
\end{align}\par
By definition, the two-point correlator $\langle{T_{zz}T_{ww}}\rangle$ can be obtained by taking the first-order functional derivative of $\braket{T_{zz}}$ with respect to $g^{(0)ww}$, namely
\begin{equation}
    \braket{T_{zz}(z)T_{ww}(w)}=\frac{-2}{\sqrt{g^{(0)}(w)}}\frac{\delta\braket{T_{zz}(z)}}{\delta g^{(0)ww}(w)}.
\end{equation}
Simultaneously, we have
\begin{equation}
    g^{(0)}_{\Bar{w}\Bar{w}}=-\frac{1}{4}e^{4\phi(w,\Bar{w})}g^{(0)ww}.
\end{equation}
Thus, the relationship between the two-point correlator and the functional derivative with respect to $\chi_{\Bar{w}\Bar{w}}$ is
\begin{equation}
    \frac{\delta\braket{T_{zz}}^{[1]}(z)}{\delta\chi_{\Bar{w}\Bar{w}}(w)}=e^{-2\phi(w,\Bar{w})}\braket{T_{zz}(z)T_{ww}(w)}\label{relation_of_der_cor}.
\end{equation}
Then we obtain
\begin{align}
    &\int_{\mathcal{D}}\dif^2z_0\,\frac{1}{\pi}\partial_{\Bar{z}_0}G^{z_0}_{\ zz}(z_0,\Bar{z}_0;z,\Bar{z})\braket{T_{zz}(z)T_{ww}(w)}=-\frac{1}{16\pi^2 G}\partial_w^3G^w_{\ zz}(w,\Bar{w};z,\Bar{z}).\label{firstpartofTT}
\end{align}\par
For the second term in (\ref{two terms in TT}), adopting the notation in\cite{Eguchi:1986sb}, we define the Teichmüller deformation of the correlator as
\begin{equation}
    \frac{1}{Z}\delta_{\mathrm{Teich}}\left(\braket{\phi_1\cdots\phi_N}_{\mathrm{tot}}Z\right)=\sum\limits_\alpha\delta\tau_\alpha\int_{\mathcal{D}}\dif^2z\,\sqrt{g^{(0)}}g^{(0)z\Bar{z}}\mu^z_{\alpha\Bar{z}}\braket{T_{zz}\phi_1\cdots\phi_N}_{\mathrm{tot}},\label{Techimuller deformation}
\end{equation}
where $\tau_\alpha$'s are modular parameters of the moduli space of the Riemann surface. We add the subscript 'tot' to distinguish these total correlators from the connected ones we consider. \par
When the curvature of the Riemann surface is a constant, we can also express the one-point correlator as an integral with holomorphic quadratic differential and Beltrami differential:
\begin{equation}
    \braket{T_{ww}}=\sum\limits_\alpha \phi_{\alpha ww}\int_{\mathcal{D}}\dif^2z\,\sqrt{g^{(0)}}g^{(0)z\Bar{z}}\mu^z_{\alpha\Bar{z}}\braket{T_{zz}}.\label{onepoint as an integral with differentials}
\end{equation}
{Combining (\ref{Techimuller deformation}) with (\ref{onepoint as an integral with differentials}),} it is straightforward to obtain
\begin{equation}
    \int_{\mathcal{D}}\dif^2z_0\,p_2(z_0,\bar{z}_0;z)\braket{T_{zz}(z_0)T_{ww}(w)}=\sum\limits_{\alpha=1}^{3g-3}\phi_{\alpha zz}\frac{\partial}{\partial \tau_\alpha}\braket{T_{ww}}.\label{secondpartofTT}
\end{equation}
This result can be generalized to the case with an $(n+1)$-point correlator on the left and an $n$-point correlator on the right.\par 
{There is also another more direct way to obtain (\ref{secondpartofTT}).} By employing the definition (\ref{definition of stress tensor correlator}) of the stress tensor correlator, we can rewrite the second term in (\ref{two terms in TT}) as follows,
\begin{align}
    \int_{\mathcal{D}}\dif^2z_0\,p_2(z_0,\bar{z}_0;z)\frac{\delta\braket{T_{zz}}^{[1]}(z_0)}{\delta\chi_{\Bar{w}\Bar{w}}(w)}&=e^{-2\phi(w,\bar w)}\int_{\mathcal{D}}\dif^2z_0\,p_2(z_0,\bar{z}_0;z)\langle{T_{z_0z_0}(z_0)T_{ww}(w)}\rangle\notag\\
    &=\sum_{\alpha=1}^{3g-3}\phi_{\alpha zz}e^{-2\phi(w,\bar w)}\int_{\mathcal{D}}\dif^2z_0\,\overline{\phi_{\alpha z_0z_0}}\frac{\delta\langle{T_{ww}(w)}\rangle}{\delta g^{(0)}_{\bar z_0\bar z_0}(z_0)}.
\end{align}
The second line describes the change in the one-point correlator $\langle{T_{ww}}\rangle$ after a physical variation of the boundary metric. As reviewed in appendix \ref{appendix Differentials and Green's function}, the physical variation of the metric is characterized by holomorphic and antiholomorphic quadratic differentials on the Riemann surface,
\begin{align}
    \delta g^{(0)}_{ij}\text{d}z^i\text{d}z^j=\sum_{\alpha=1}^{3g-3}\Big(\phi_{\alpha zz}\delta\bar\tau_{\alpha}(\text{d}z)^2+\overline{\phi_{\alpha zz}}\delta\tau_{\alpha}(\text{d}\bar z)^2\Big).
\end{align}
Then it is straightforward to obtain
\begin{align}
    \int_{\mathcal{D}}\dif^2z_0\,\overline{\phi_{\alpha z_0z_0}}\frac{\delta\langle{T_{ww}(w)}\rangle}{\delta g^{(0)}_{\bar z_0\bar z_0}(z_0)}=\frac{\partial}{\partial\tau_{\alpha}}\langle{T_{ww}(w)}\rangle,
\end{align}
{which immediately reproduces (\ref{secondpartofTT}).}\par
Combine (\ref{firstpartofTT}) with (\ref{secondpartofTT}), we finally obtain the result
\begin{equation}
    \braket{T_{zz}(z)T_{ww}(w)}=-\frac{1}{16\pi^2 G}\partial_w^3G^w_{\ zz}(w,\Bar{w};z,\Bar{z})+\sum\limits_{\alpha=1}^{3g-3}\phi_{\alpha zz}\frac{\partial}{\partial \tau_\alpha}\braket{T_{ww}(w)},
\end{equation}
which matches the result from Ward identity in \cite{Eguchi:1986sb} correctly. The specific expressions of the Green's function $G^w_{\ zz}$, the holomorphic quadratic differentials $\phi_{\alpha zz}$, and modular parameters $\tau_\alpha$ depend on the genus of the Riemann surface and the basis we choose.\par
Two-point correlators of other stress tensor components can be obtained by the same method. Here we list all independent two-point correlators:
\begin{align}
    \braket{T_{zz}(z)T_{ww}(w)}=&-\frac{1}{16\pi^2 G}\partial_w^3G^w_{\ zz}(w,\Bar{w};z,\Bar{z})+\sum\limits_{\alpha=1}^{3g-3}\phi_{\alpha zz}\frac{\partial}{\partial \tau_\alpha}\braket{T_{ww}},\\
    \braket{T_{zz}(z)T_{\Bar{w}\Bar{w}}(w)}=&\ \frac{1}{16\pi G}(4\partial_w\phi\partial_{\Bar{w}}\phi-\partial_w\phi\partial_{\Bar{w}}+2\partial_{\Bar{w}}\phi\partial_w-8\partial_w\partial_{\Bar{w}}\phi-\partial_w\partial_{\Bar{w}})\delta^{(2)}(w-z)\nonumber\\
    &+\frac{3}{8\pi G}\partial_w\partial_{\Bar{w}}\phi \,p_2(w,\bar{w};z)+\sum\limits_{\alpha=1}^{3g-3}\phi_{\alpha zz}\frac{\partial}{\partial \tau_\alpha}\braket{T_{\Bar{w}\Bar{w}}},\\
    \braket{T_{z\Bar{z}}(z)T_{ww}(w)}=&\ \frac{1}{16\pi G}\big(2\partial_z^2\phi-2(\partial_z\phi)^2-2\partial_z\phi\partial_z+\partial_z^2\big)\delta^{(2)}(z-w),\\
    \braket{T_{z\Bar{z}}(z)T_{w\Bar{w}}(w)}=&\ \frac{1}{16\pi G}\big(2\partial_z\phi\partial_{\Bar{z}}+2\partial_{\Bar{z}}\phi\partial_z-4\partial_z\phi\partial_{\Bar{z}}\phi-\partial_z\partial_{\Bar{z}}\big)\delta^{(2)}(z-w).
\end{align}

\subsection{Recurrence relations and higher-point correlators}\label{subsection Recurrence relations and higher-point correlators}
We can also compute higher point correlators and derive valuable recurrence relations concerning them. By taking the $n$-th ($n\geqslant2$) functional derivative of the $n$-th order of (\ref{conservlaw}) with respect to $\chi_{\Bar{z}\Bar{z}}$ and evaluating the result in the unperturbed metric, we obtain
\begin{align}
    &\partial_{\Bar{z}}\frac{\delta^n\braket{T_{zz}}^{[n]}(z)}{\prod\limits_{i=1}^n\delta\chi_{\Bar{z}\Bar{z}}(z_i)}-e^{-2\phi(z,\Bar{z})}\sum\limits_{i=1}^n\partial_z\frac{\delta^{n-1}\braket{T_{zz}}^{[n-1]}(z)}{\prod\limits_{j\neq i}\delta\chi_{\Bar{z}\Bar{z}}(z_j)}\delta^{(2)}(z-z_i)\nonumber\\
    &\quad-2e^{-2\phi(z,\Bar{z})}\sum\limits_{i=1}^n\frac{\delta^{n-1}\braket{T_{zz}}^{[n-1]}(z)}{\prod\limits_{j\neq i}\delta\chi_{\Bar{z}\Bar{z}}(z_j)}\partial_z\delta^{(2)}(z-z_i)\nonumber\\
    &\quad+4e^{-2\phi(z,\Bar{z})}\sum\limits_{i=1}^n\frac{\delta^{n-1}\braket{T_{zz}}^{[n-1]}(z)}{\prod\limits_{j\neq i}\delta\chi_{\Bar{z}\Bar{z}}(z_j)}\partial_z\phi\delta^{(2)}(z-z_i)=0\label{preTTrecur}.
\end{align}
Following (\ref{relation_of_der_cor}), we have
\begin{equation}
    \frac{\delta^n\braket{T_{zz}}^{[n]}(z)}{\delta\chi_{\Bar{z}\Bar{z}}(z_1)\cdots\delta\chi_{\Bar{z}\Bar{z}}(z_n)}=e^{-2\phi(z_1,\Bar{z}_1)}\cdots e^{-2\phi(z_n,\Bar{z}_n)}\braket{T_{zz}(z)T_{zz}(z_1)\cdots T_{zz}(z_n)},
\end{equation}
thus (\ref{preTTrecur}) becomes
\begin{align}
    &e^{2\phi(z,\Bar{z})}\partial_{\Bar{z}}\braket{T_{zz}(z)T_{zz}(z_1)\cdots T_{zz}(z_n)}\nonumber\\
    &=\sum\limits_{i=1}^n e^{2\phi(z_i,\Bar{z}_i)}\partial_z\braket{T_{zz}(z)T_{zz}(z_1)\cdots T_{zz}(z_{i-1})T_{zz}(z_{i+1})\cdots T_{zz}(z_n)}\delta^{(2)}(z-z_i)\nonumber\\
    &\quad\,+2\sum\limits_{i=1}^n e^{2\phi(z_i,\Bar{z}_i)}\braket{T_{zz}(z)T_{zz}(z_1)\cdots T_{zz}(z_{i-1})T_{zz}(z_{i+1})\cdots T_{zz}(z_n)}\partial_z\delta^{(2)}(z-z_i)\nonumber\\
    &\quad\,-4\sum\limits_{i=1}^n e^{2\phi(z_i,\Bar{z}_i)}\braket{T_{zz}(z)T_{zz}(z_1)\cdots T_{zz}(z_{i-1})T_{zz}(z_{i+1})\cdots T_{zz}(z_n)}\partial_z\phi(z,\Bar{z})\delta^{(2)}(z-z_i).
\end{align}
Solved with Green's function, we have
\begin{align}
    &\braket{T_{zz}(z)T_{zz}(z_1)\cdots T_{zz}(z_n)}=\frac{2}{\pi}\sum\limits_{i=1}^n\partial_{z_i}G^{z_i}_{\ zz}(z_i,\Bar{z}_i;z,\Bar{z})\braket{T_{zz}(z_1)\cdots T_{zz}(z_n)}\nonumber\\
    &\ \,+\frac{1}{\pi}\sum\limits_{i=1}^n G^{z_i}_{\ zz}(z_i,\Bar{z}_i;z,\Bar{z})\partial_{z_i}\braket{T_{zz}(z_1)\cdots T_{zz}(z_n)}+\sum\limits_{\alpha=1}^{3g-3}\phi_{\alpha zz}(z)\frac{\partial}{\partial \tau_\alpha}\braket{T_{zz}(z_1)\cdots T_{zz}(z_n)},
\end{align}
which gives the recurrence relation of correlators of $T_{zz}$ components.\par
In particular, we can obtain the three-point correlators $\braket{T_{zz}T_{zz}T_{zz}}$ directly from the recurrence relation above:
\begin{align}
    &\braket{T_{zz}(z)T_{zz}(z_1)T_{zz}(z_2)}=\frac{2}{\pi}\sum\limits_{i=1}^2\partial_{z_i}G^{z_i}_{\ zz}(z_i,\Bar{z}_i;z,\Bar{z})\braket{T_{zz}(z_1)T_{zz}(z_2)}\nonumber\\
    &\quad\quad+\frac{1}{\pi}\sum\limits_{i=1}^2 G^{z_i}_{\ zz}(z_i,\Bar{z}_i;z,\Bar{z})\partial_{z_i}\braket{T_{zz}(z_1)T_{zz}(z_2)}+\sum\limits_{\alpha=1}^{3g-3}\phi_{\alpha zz}(z)\frac{\partial}{\partial \tau_\alpha}\braket{T_{zz}(z_1)T_{zz}(z_2)}.
\end{align}\par
There are also other useful relations. For example, by taking the $n$-th ($n\geqslant3$) functional derivative of the $n$-th order of (\ref{conservlaw}) with respect to $\chi_{zz}$ and evaluating the result in the unperturbed metric, we have
\begin{align}
    &\partial_{\Bar{z}}\frac{\delta^n\braket{T_{zz}}^{[n]}(z)}{\delta\chi_{zz}(z_1)\cdots\delta\chi_{zz}(z_n)}+e^{-2\phi(z,\Bar{z})}\sum\limits_{i=1}^n\delta^{(2)}(z-z_i)\partial_{z}\frac{\delta^{n-1}\braket{T_{\Bar{z}\Bar{z}}}^{[n-1]}(z)}{\delta\chi_{zz}(z_1)\cdots\delta\chi_{zz}(z_{i-1})\delta\chi_{zz}(z_{i+1})\cdots\delta\chi_{zz}(z_n)}\nonumber\\
    &-2e^{-4\phi(z,\Bar{z})}\frac{1}{(n-2)!}\sum\limits_{\sigma\in S_n}\bigg(2\delta^{(2)}(z-z_{\sigma(1)})\frac{\delta^{n-2}\braket{T_{\Bar{z}\Bar{z}}}^{[n-2]}(z)}{\delta\chi_{zz}(z_{\sigma(2)})\cdots\delta\chi_{zz}(z_{\sigma(n-1)})}\partial_{\Bar{z}}\delta^{(2)}(z-z_{\sigma(n)})\nonumber\\
    &\ +\delta^{(2)}(z-z_{\sigma(1)})\delta^{(2)}(z-z_{\sigma(2)})\partial_{\Bar{z}}\frac{\delta^{n-2}\braket{T_{\Bar{z}\Bar{z}}}^{[n-2]}(z)}{\delta\chi_{zz}(z_{\sigma(3)})\cdots\delta\chi_{zz}(z_{\sigma(n)})}\nonumber\\
    &\ +4\delta^{(2)}(z-z_{\sigma(1)})\delta^{(2)}(z-z_{\sigma(2)})\frac{\delta^{n-2}\braket{T_{\Bar{z}\Bar{z}}}^{[n-2]}(z)}{\delta\chi_{zz}(z_{\sigma(3)})\cdots\delta\chi_{zz}(z_{\sigma(n)})}\partial_{\Bar{z}}\phi(z,\Bar{z})\bigg)=0.\label{norderttbartbar}
\end{align}
Solved with Green's function, finally, we have
\begin{align}
    &\braket{T_{zz}(z)T_{\Bar{z}\Bar{z}}(z_1)\cdots T_{\Bar{z}\Bar{z}}(z_n)}=\frac{1}{\pi}\sum\limits_{i=1}^nG^{z_i}_{\ zz}(z_i,\Bar{z}_i;z,\Bar{z})\partial_{\Bar{z}_i}\braket{T_{\Bar{z}\Bar{z}}(z_1)\cdots T_{\Bar{z}\Bar{z}}(z_n)}\nonumber\\
    &-\frac{2}{(n-2)!\pi}\sum\limits_{\sigma\in S_n}G^{z_{\sigma(1)}}_{\ zz}(z_{\sigma(1)},\Bar{z}_{\sigma(1)};z,\Bar{z})\Big(2\braket{T_{\Bar{z}\Bar{z}}(z_{\sigma(1)})\cdots T_{\Bar{z}\Bar{z}}(z_{\sigma(n-1)})}\partial_{\Bar{z}_{\sigma(1)}}\nonumber\\
    &\ +\partial_{\Bar{z}_{\sigma(1)}}\braket{T_{\Bar{z}\Bar{z}}(z_{\sigma(1)})\cdots T_{\Bar{z}\Bar{z}}(z_{\sigma(n-1)})}+4\braket{T_{\Bar{z}\Bar{z}}(z_{\sigma(1)})\cdots T_{\Bar{z}\Bar{z}}(z_{\sigma(n-1)})}\partial_{\Bar{z}_{\sigma(1)}}\phi\Big)\delta^{(2)}(z_{\sigma(1)}-z_{\sigma(n)})\nonumber\\
    &\ +\sum\limits_{\alpha=1}^{3g-3}\phi_{\alpha zz}(z)\frac{\partial}{\partial \tau_\alpha}\braket{T_{\Bar{z}\Bar{z}}(z_1)\cdots T_{\Bar{z}\Bar{z}}(z_n)},
\end{align}
which gives the relation between the correlator of one $T_{zz}$ component, $n$ $T_{\Bar{z}\Bar{z}}$ components and the correlator of $n$ $T_{\Bar{z}\Bar{z}}$ components ($n\geqslant3$).\par
We can also obtain the relation between the correlator of one $T_{z\Bar{z}}$ component, $n$ $T_{zz}$ components, and the correlator of $n$ $T_{zz}$ components $(n\geqslant2)$. Taking the $n$-th derivative of the $n$-th order of (\ref{confanomaly}) with respect to $\chi_{\Bar{z}\Bar{z}}$ {and} evaluating the result in the unperturbed metric, we have
\begin{align}
    &\braket{T_{z\Bar{z}}(z)T_{zz}(z_1)\cdots T_{zz}(z_n)}\nonumber\\
    &\qquad\qquad\qquad=\sum\limits_{i=1}^n\braket{T_{zz}(z)T_{zz}(z_1)\cdots T_{zz}(z_{i-1})T_{zz}(z_{i+1})\cdots T_{zz}(z_n)}\delta^{(2)}(z-z_i).
\end{align}\par
Applying the same method, we compute all independent three-point and four-point correlators of the stress tensor. However, many contain numerous contact terms, so we leave them to appendix \ref{appendix List of three-point and four-point correlators}.

\section{Holographic correlators at finite cutoff}\label{section Holographic correlators at finite cutoff}
In the previous sections, we computed the holographic stress tensor correlators on a general Riemann surface within the framework of AdS$_3$/CFT$_2$. In what follows, we will investigate the holographic aspects of a cutoff AdS$_3$. Let $(\rho,z,\bar z)$ be the FG coordinates in the bulk. The Dirichlet boundary $\partial\mathcal{M}_c$ is a hard radial cutoff at $\rho=\rho_c$. The generalized GKPW relation gives a natural holographic dictionary for cutoff AdS$_3$,
\begin{align}
    Z_{G}[g^{(c)}_{ij}]=\Big\langle{\text{exp}{\Big[-\frac{1}{2}\int\text{d}^2z\sqrt{g^{(c)}}g^{(c)ij}T_{ij}\Big]}}\Big\rangle_{\text{EFT}}, \label{generalized GKPW relation}
\end{align}
where the sources $g^{(c)}_{ij}(z,\bar z)=g_{ij}(\rho_c,z,\bar z)$ are the components of the boundary metric. The dual EFT on the right-hand side is obtained by $T\bar T$ deformation of the original CFT \cite{McGough:2016lol}, which is defined by the following flow equation for the action,
\begin{align}
    \frac{\text{d}S_{\lambda}}{\text{d}\lambda}=-\frac{1}{4}\int\text{d}^2z\,\text{det}[T_{\lambda}].
\end{align}
The deformation parameter $\lambda$ is related to the cutoff location $\rho_c$ by 
\begin{align}
    \lambda=16\pi G \rho_c.
\end{align}\par
This paper is concerned with the holographic stress tensor correlators in cutoff AdS$_3$. The one-point correlator is identified with the Brown-York tensor \cite{Brown:1992br} evaluated on the cutoff surface,
\begin{align}
    \langle{T_{ij}}\rangle_{\rho_c}=&-\frac{1}{8\pi G}(K^{(c)}_{ij}-K^{(c)}h^{(c)}_{ij}+h^{(c)}_{ij}).\label{TTbar deformed one-point correlator}
\end{align}
Plugging (\ref{TTbar deformed one-point correlator}) into Einstein's equation to replace the extrinsic curvature, we obtain
\begin{align}
&\nabla^{i}\langle{T_{ij}}\rangle_{\rho_c}=0,\label{TTbar deformed conservation equation}\\
&\langle{T^{i}_{i}}\rangle_{\rho_c}=\frac{1}{16\pi G}R^{(c)}-8\pi G\rho_c\text{det}[T]_{\rho_c},\label{TTbar deformed trace relation}\\
&\partial_{\rho_c}\langle{T_{ij}}\rangle_{\rho_c}=4\pi G[2\langle{T^{k}_{i}}\rangle_{\rho_c}\langle{T_{kj}}\rangle_{\rho_c}-\langle{T^{k}_{k}}\rangle_{\rho_c}\langle{T_{ij}}\rangle_{\rho_c}-\text{det}[T]_{\rho_c}g^{(c)}_{ij}],  \label{TTbar deformed radial flow equation}
\end{align}
where the indices are raised by $g^{(c)ij}$ and $\text{det}[T]_{\rho_c}=\frac{1}{2}(\langle{T_{i}^{i}}\rangle_{\rho_c}^2-\langle{T^{ij}}\rangle_{\rho_c}\langle{T_{ij}}\rangle_{\rho_c})$. The first two equations (\ref{TTbar deformed conservation equation}) and (\ref{TTbar deformed trace relation}) represent the conservation equation and the trace relation of the stress tensor, respectively. These equations are subsequently utilized for calculating two-point correlators. The final equation (\ref{TTbar deformed radial flow equation}), which characterizes the radial flow effect of the stress tensor within the same FG coordinate system, will be employed to compute the deformed one-point correlators.

\subsection{Dynamical coordinate transformation}\label{subsection Dynamical coordinate transformation}
The gravitational partition function on the left-hand side of (\ref{generalized GKPW relation}) can be approximated as a sum over all saddles in the semiclassical limit, with the dominant saddle being assumed to be the handlebody solution. In contrast to the solution employed in section \ref{section Holographic correlators of higher genus CFT}, here we fix the metric at a finite cutoff instead of on the conformal boundary.
The boundary metrics on various radial slices can be related by employing the dynamic coordinate transformation \cite{Dubovsky:2012wk,Dubovsky:2017cnj,Conti:2018tca} and Weyl transformation \cite{Caputa:2020lpa, Tian:2023fgf}. \par
In the following, we derive the explicit form of the dynamical coordinate transformation for a Riemann surface as the cutoff boundary. In a certain FG coordinate system, the metric on a given fixed $\rho$ slice is expressed in the conformal gauge as
\begin{align}
    g_{ij}(\rho,z,\bar z)\text{d}x^i\text{d}x^j=e^{2\omega_{\rho}(z,\bar z)}\text{d}z\text{d}\bar z.
\end{align}
Meanwhile, the Riemann surface at $\rho$ is constructed by taking the quotient of $\mathbb{C}\cup\lbrace{\infty}\rbrace\backslash \Lambda(\Gamma_{\rho})$ with respect to some Schottky group $\Gamma_{\rho}$. To ensure the invariance of the line element under the action of $\Gamma_{\rho}$, it is necessary for the Weyl factor $\omega_{\rho}$ to satisfy the following equivariance condition,
\begin{align}
    \omega_{\rho}(\gamma_{\rho}(z),\overline{\gamma_{\rho}(z)})=\omega_{\rho}(z,\bar z)-\frac{1}{2}\text{ln}|\gamma_{\rho}'(z)|^2, \label{equivariance condition 1}
\end{align}
for any $\gamma_{\rho}\in\Gamma_{\rho}$. Consider a small radial shift $\delta\rho$ of the cutoff boundary. The variation of the boundary metric can be expressed in the original FG coordinate system as follows:
\begin{align}
    \delta_\rho g_{ij}=(g^{(2)}_{ij}+2\rho g^{(4)}_{ij})\delta\rho=8\pi G(\langle{T_{ij}}\rangle_{\rho}-g^{kl}\langle{T_{kl}}\rangle_{\rho}g_{ij})\delta\rho.
\end{align}
Under a tangential coordinate transformation $\delta x^i=\epsilon_{\rho}^i$, the metric on the new boundary (i.e. the hard radial cutoff at $\rho+\delta\rho$) can be rewritten in the conformal gauge, and the metric variation corresponds to an infinitesimal Weyl transformation,
\begin{align}
    \delta g_{ij}=\delta_\rho g_{ij}+\mathcal{L}_{\epsilon}g_{ij}=2\delta\omega_{\rho}g_{ij}.
\end{align}
It follows that
\begin{align}
    &\partial_{\bar z}\epsilon_{\rho}^{z}=-8\pi G e^{-2\omega_{\rho}}\langle{T_{\bar z\bar z}}\rangle_{\rho}\delta\rho,\ \ \ \partial_z\epsilon_{\rho}^{\bar z}=-8\pi G e^{-2\omega_{\rho}}\langle{T_{zz}}\rangle_{\rho}\delta\rho,\label{DE of diffeomorphism}\\
    &\delta\omega_{\rho}=\frac{1}{2}e^{-2\omega_{\rho}}[\partial_z(e^{2\omega_{\rho}}\epsilon_{\rho}^z)+\partial_{\bar z}(e^{2\omega_{\rho}}\epsilon_{\rho}^{\bar z})-16\pi G\langle{T_{z\bar z}}\rangle_{\rho}\delta\rho]\label{variation of Weyl Weyl factor}.
\end{align}
The variation of the stress tensor one-point correlator is also divided into two parts: one arises from the radial flow in the original FG coordinate system (\ref{TTbar deformed radial flow equation}), and the other originates from the tangential coordinate transformation,
\begin{align}
\mathcal{L}_\epsilon\langle{T_{ij}}\rangle_{\rho}=\epsilon_{\rho}^k\partial_k\langle{T_{ij}}\rangle_{\rho}+\partial_i\epsilon_{\rho}^k\langle{T_{kj}}\rangle_{\rho}+\partial_j\epsilon_{\rho}^k\langle{T_{ik}}\rangle_{\rho}. \label{variation of 1-pt from diffeomorphism}
\end{align}
Combining (\ref{TTbar deformed radial flow equation})(\ref{DE of diffeomorphism})(\ref{variation of 1-pt from diffeomorphism}) we obtain
\begin{align}
    \delta\langle{T_{zz}}\rangle_{\rho}&=(\epsilon^k\partial_k+2\partial_z\epsilon^z)\langle{T_{zz}}\rangle_{\rho},\notag\\
    \delta\langle{T_{\bar z\bar z}}\rangle_{\rho}&=(\epsilon^k\partial_k+2\partial_{\bar z}\epsilon^{\bar z})\langle{T_{zz}}\rangle_{\rho},\notag\\
    \delta\langle{T_{z\bar z}}\rangle_{\rho}&=(\epsilon^k\partial_k+\partial_k\epsilon^k)\langle{T_{z\bar z}}\rangle_{\rho}-2\pi G e^{2\omega_{\rho}}\text{det}[T]_{\rho}\delta\rho.\label{variation of 1-pt}
\end{align}
Next, we need to find the explicit form of the diffeomorphism $\epsilon_{\rho}^i$ that satisfies (\ref{DE of diffeomorphism}). In \cite{Caputa:2020lpa}, the authors present a construction of $\epsilon^i$ on a curved plane.
However, directly extending this construction to a Riemann surface can't be feasible, as the metric $e^{2\omega}$ and the stress tensor correlators $\langle{T_{z\bar z}}\rangle$, $\langle{T_{zz}}\rangle$ and $\langle{T_{\bar z\bar z}}\rangle$ on a Riemann surface should exhibit ``periodicity''. In the Schottky uniformization, this ``periodicity'' implies that $e^{2\omega}$, $\langle{T_{z\bar z}}\rangle$, $\langle{T_{zz}}\rangle$, and $\langle{T_{\bar z\bar z}}\rangle$ are automorphic forms of type $(1,1)$, $(1,1)$, $(2,0)$, and $(0,2)$ respectively. As we vary the radial coordinate of the boundary, the Schottky group associated with the Riemann surface also changes, which is described by a curve in the modular space,
\begin{align}
    \tau_{\alpha}=\tau_{\alpha}(\rho),\ \ \ \alpha=1,2,...,3g-3.
\end{align}
Assuming that $e^{2\omega_{\rho}}$ and $\langle{T_{ij}}\rangle_\rho$ are already automorphic forms with respect to the Schottky group $\Gamma_{\rho}$. After a small radial shift $\delta\rho$, the metric and stress tensor correlators on the new boundary should be manifested as the automorphic forms to another Schottky group $\Gamma_{\rho+\delta\rho}$, i.e. satisfy
 \begin{align}
     e^{2\omega_{\rho+\delta\rho}(z,\bar z)}&=\gamma'_{\rho+\delta\rho}(z)\overline{\gamma'_{\rho+\delta\rho}(z)}e^{2\omega_{\rho+\delta\rho}(\gamma_{\rho+\delta\rho}(z),\overline{\gamma_{\rho+\delta\rho}(z)})},\notag\\
     \langle{T_{z\bar z}(z,\bar z)}\rangle_{\rho+\delta\rho}&=\gamma'_{\rho+\delta\rho}(z)\overline{\gamma'_{\rho+\delta\rho}(z)}\langle{T_{z\bar z}(\gamma_{\rho+\delta\rho}(z),\overline{\gamma_{\rho+\delta\rho}(z)})}\rangle_{\rho+\delta\rho},\notag\\
          \langle{T_{zz}(z,\bar z)}\rangle_{\rho+\delta\rho}&=(\gamma'_{\rho+\delta\rho}(z))^2\langle{T_{zz}(\gamma_{\rho+\delta\rho}(z),\overline{\gamma_{\rho+\delta\rho}(z)})}\rangle_{\rho+\delta\rho},\notag\\
               \langle{T_{\bar z\bar z}(z,\bar z)}\rangle_{\rho+\delta\rho}&=(\overline{\gamma'_{\rho+\delta\rho}(z)})^2\langle{T_{\bar z\bar z}(\gamma_{\rho+\delta\rho}(z),\overline{\gamma_{\rho+\delta\rho}(z)})}\rangle_{\rho+\delta\rho}, \label{periodicity condition for metric and 1-pt}
 \end{align}
where $\gamma'_{\rho+\delta\rho}(z)=\frac{\text{d}\gamma_{\rho+\delta\rho}(z)}{\text{d}z}$. The variations of the Weyl factor $\delta\omega_\rho$ and the one-point correlators $\delta\langle{T_{ij}}\rangle_{\rho}$ are determined by (\ref{variation of Weyl Weyl factor}) and (\ref{variation of 1-pt}), respectively. By combining these with (\ref{periodicity condition for metric and 1-pt}), we can deduce the periodicity conditions of $\epsilon^{z}_{\rho}$ and $\epsilon^{\bar z}_{\rho}$,
\begin{align}
    \epsilon_{\rho}^{z}(\gamma_{\rho}(z),\overline{\gamma_{\rho}(z)})&=\gamma'_{\rho}(z)\epsilon_{\rho}^{z}(z,\bar z)-\delta\gamma_{\rho}(z),\notag\\
    \epsilon_{\rho}^{\bar z}(\gamma_{\rho}(z),\overline{\gamma_{\rho}(z)})&=\overline{\gamma'_{\rho}(z)}\epsilon_{\rho}^{\bar z}(z,\bar z)-\delta\overline{\gamma_{\rho}(z)}. \label{periodicity condition of epsilon}
\end{align}
One can observe that $\epsilon^{z}_{\rho}$ is not an automorphic form. It is multiple-valued on the Riemann surface, and the discontinuity corresponds to the variation of the Schottky group element $\gamma_{\rho}$. Based on (\ref{DE of diffeomorphism}) and (\ref{periodicity condition of epsilon}), we present the following construction,
\begin{align}
    \epsilon_{\rho}^z&=-8G\delta\rho\int_{\mathcal{D}_{\rho}}e^{-2\omega_{\rho}(w,\bar w)}\langle{T_{\bar w\bar w}}\rangle_{\rho}\Big[G^{z}_{ww}+\sum_{\alpha=1}^{3g-3}f^{z}_{\alpha}\phi_{\alpha ww}\Big]_{\Gamma_{\rho}}\text{d}^2w,\notag\\
        \epsilon_{\rho}^{\bar z}&=-8G\delta\rho\int_{\mathcal{D}_{\rho}} 
        e^{-2\omega_{\rho}(w,\bar w)}\langle{T_{ww}}\rangle_{\rho}\Big[\overline{G^{z}_{ww}}+\sum_{\alpha=1}^{3g-3}\overline{f^{z}_{\alpha}} \overline{\phi_{\alpha ww}}\Big]_{\Gamma_{\rho}}\text{d}^2w,\label{diffeomorphism on the Riemann surface}
\end{align}
where $\mathcal{D}_{\rho}$ is the fundamental domain for $\Gamma_{\rho}$. $G^{z}_{ww}$ is the Green's function on the Riemann surface, and $\lbrace{\phi_{\alpha ww}}\rbrace$ forms a basis of the space $\mathcal{H}_g^2$, both of which have been employed in section \ref{section Holographic correlators of higher genus CFT}. $f_{\alpha}^{z}$ is the Bers potential (as defined in appendix \ref{appendix Differentials and Green's function}), which is associated with the Beltrami differential through
\begin{align}
    \frac{1}{\pi}\partial_{\bar z}f_{\alpha}^{z}=\mu_{\alpha\bar z}^{z},\ \ \ \alpha=1,2,...,3g-3.\label{differential equation for Bers potential}
\end{align}
$f_{\alpha}^{z}$ is not an automorphic form, and its discontinuity can be written as
\begin{align}
    f_{\alpha}^{z}(\gamma_{\rho}(z),\overline{\gamma_{\rho}(z)})-\gamma'_{\rho}(z)f_{\alpha}^{z}(z,\bar z)=\gamma'_{\rho}(z)\Xi_{\alpha}^{z}[\gamma_{\rho}](z).
\end{align}
The function\footnote{Moreover, $\Xi_{\alpha}^{z}[\gamma_{\rho}](z)$ is the component of an Eichler $1$-cocycle for $\Gamma_{\rho}$.} $\Xi_{\alpha}^{z}[\gamma_{\rho}](z)$ is polynomial in $z$ of degree $2$ \cite{Tuite:2019fqx}. The discontinuity of Bers potential governs the flow of the corresponding Schottky group element in the modular space \cite{Tuite:2019fqx,Roland:1993pm,Playle:2015sxa},
\begin{align}
    \frac{\partial\gamma_{\rho}(z)}{\partial\tau_{\alpha}}=\frac{1}{\pi}\gamma'_{\rho}(z)\Xi_{\alpha}^{z}[\gamma_{\rho}](z),\ \ \ \alpha=1,2,...,3g-3.
\end{align}
Returning to the construction of $\epsilon_{\rho}^{i}$ in (\ref{diffeomorphism on the Riemann surface}), it becomes apparent, based on the definitions in (\ref{differential equation for G}) and (\ref{differential equation for Bers potential}), that $\epsilon_{\rho}^i$ satisfies the differential equation (\ref{DE of diffeomorphism}). Moreover, the discontinuities of $\epsilon_{\rho}^{z}$ and $\epsilon_{\rho}^{\bar z}$ are given by
\begin{align}
    \epsilon_{\rho}^{z}(\gamma_{\rho}(z),\overline{\gamma_{\rho}(z)})-\gamma'_{\rho}(z)\epsilon_{\rho}^{z}(z,\bar z)=&-8\pi G\delta\rho\sum_{\alpha=1}^{3g-3}\Big[\frac{\partial\gamma_{\rho}(z)}{\partial\tau_{\alpha}}\int_{\mathcal{D}_{\rho}}e^{-2\omega_{\rho}(w,\bar w)}\langle{T_{\bar w\bar w}}\rangle_{\rho}\phi_{\alpha ww}\text{d}^2w\Big],\notag\\
    \epsilon_{\rho}^{\bar z}(\gamma_{\rho}(z),\overline{\gamma_{\rho}(z)})-\overline{\gamma'_{\rho}(z)}\epsilon_{\rho}^{\bar z}(z,\bar z)=&-8\pi G\delta\rho\sum_{\alpha=1}^{3g-3}\Big[\frac{\partial\overline{\gamma_{\rho}(z)}}{\partial\bar\tau_{\alpha}}\int_{\mathcal{D}_{\rho}}e^{-2\omega_{\rho}(w,\bar w)}\langle{T_{ww}}\rangle_{\rho}\overline{\phi_{\alpha ww}}\text{d}^2w\Big].\label{periodicity condition of epsilon 2}
\end{align}
Comparing equations (\ref{periodicity condition of epsilon}) and (\ref{periodicity condition of epsilon 2}), since $\delta\gamma_{\rho}=\delta\rho\sum_{\alpha=1}^{3g-3}\frac{\partial\gamma_{\rho}}{\partial\tau_{\alpha}}\frac{\text{d}\tau_{\alpha}}{\text{d}\rho}$, we read off
\begin{align}
    \frac{\text{d}\tau_{\alpha}}{\text{d}\rho}=8\pi G\int_{\mathcal{D}_{\rho}}e^{-2\omega_{\rho}(w,\bar w)}\langle{T_{\bar w\bar w}}\rangle_{\rho}\phi_{\alpha ww}\text{d}^2w,\ \ \ \alpha=1,2,...,3g-3. \label{dynamical modular flow}
\end{align}
One can observe that the changes in the modular parameters are also dynamic since they depend on the deformed stress tensor correlators $\langle{T_{ww}}\rangle_\rho$ and $\langle{T_{\bar w\bar w}}\rangle_\rho$. In the perturbative calculation, the dynamical flows (\ref{dynamical modular flow}) of the modular parameters do not contribute to the first-order correction of the correlator (as we will demonstrate in the next subsection); however, they play a crucial role in computing higher-order corrections of the correlator.
\subsection{Perturbative stress tensor one-point correlator}\label{subsection Perturbative stress tensor one-point correlator}
Starting from AdS$_3$ with the boundary located at $\rho=0$, the boundary metric is written in the conformal gauge as $g_{ij}\text{d}x^i\text{d}x^j=e^{2\omega_{0}}\text{d}z\text{d}\bar z$. Following the approach in \cite{Krasnov:2001cu, Imbimbo:1999bj, Skenderis:2000in}, the near-boundary solution is constructed by transforming Poincar\'e AdS$_3$ using a bulk diffeomorphism that preserves the FG gauge, and the stress tensor one-point correlator takes the form
\begin{align}
        \langle{T_{ij}}\rangle_{\rho=0}=\frac{1}{8\pi G}(\partial_i\partial_j\omega_0-\partial_i\omega_0\partial_j\omega_0-\eta^{kl}\partial_k\partial_l\omega_0\eta_{ij}+\frac{1}{2}\eta^{kl}\partial_{k}\omega_0\partial_l\omega_0\eta_{ij}).\label{handlebody solution stress tensor 1pt 1}
\end{align}
Since $\omega_0$ satisfies the equivariance condition (\ref{equivariance condition 1}), it is easy to check that the components of $\langle{T_{ij}}\rangle_{\rho=0}$ are automorphic forms. Then, we need to solve the coupled nonlinear equations (\ref{variation of Weyl Weyl factor}) and (\ref{variation of 1-pt}) by employing the diffeomorphism (\ref{diffeomorphism on the Riemann surface}), while imposing Dirichlet boundary conditions at $\rho=\rho_c$,
\begin{align}
   \omega_{\rho_c}(z,\bar z)=\phi(z,\bar z),\ \ \ \Gamma_{\rho_c}=\Gamma.
\end{align}
In general, obtaining the exact solution can be quite challenging; however, the perturbation method remains viable. Expanding $\omega_{0}$ and $\langle{T_{ij}}\rangle_{\rho_c}$ in $\rho_c$,
\begin{align}
    \omega_{0}(z,\bar z)=\phi(z,\bar z)+\sum_{n=1}^{\infty}\rho_c^n\phi_n(z,\bar z),\ \ \ \langle{T_{ij}(z,\bar z)}\rangle_{\rho_c}=\sum_{n=0}^{\infty}\rho_c^n\langle{T_{ij}(z,\bar z)}\rangle_{n},
\end{align}
and plugging them into (\ref{variation of Weyl Weyl factor}) and (\ref{variation of 1-pt}). At the leading order, $\langle{T_{ij}}\rangle_0$ agrees with the CFT one-point correlator. At the subleading order, we obtain
\begin{align}
    \phi_{1}(z,\bar z)&=\frac{1}{2\pi}\Big[-\frac{\pi}{4}+\int_{\mathcal{D}}e^{-2\phi(w,\bar w)}\Big((\partial_{\bar w}^2\phi-(\partial_{\bar w}\phi)^2)(\partial_z+2\partial_z\phi)(G^z_{ww}+\sum_{\alpha=1}^{3g-3}f_{\alpha}^{z}\phi_{\alpha ww})\notag\\
    &\quad+(\partial_{w}^2\phi-(\partial_{w}\phi)^2)(\partial_{\bar z}+2\partial_{\bar z}\phi)(\overline{G^z_{ww}}+\sum_{\alpha=1}^{3g-3}\overline{f_{\alpha}^{z}}\overline{\phi_{\alpha ww}})\Big)\text{d}^2w\Big],\notag\\
    \langle{T_{zz}}\rangle_1&=\frac{1}{16\pi^2G}\Big[\frac{\pi}{4}(\partial_z^2\phi-(\partial_z\phi)^2)+\int_{\mathcal{D}}e^{-2\phi(w,\bar w)}(\partial_{\bar w}^2\phi-(\partial_{\bar w}\phi)^2)\partial_z^3(G^z_{ww}+\sum_{\alpha=1}^{3g-3}f_{\alpha}^{z}\phi_{\alpha ww})\text{d}^2w\Big],\notag\\
        \langle{T_{\bar z\bar z}}\rangle_1&=\frac{1}{16\pi^2G}\Big[\frac{\pi}{4}(\partial_{\bar z}^2\phi-(\partial_{\bar z}\phi)^2)+\int_{\mathcal{D}}e^{-2\phi(w,\bar w)}(\partial_{w}^2\phi-(\partial_{w}\phi)^2)\partial_{\bar z}^3(\overline{G^z_{ww}}+\sum_{\alpha=1}^{3g-3}\overline{f_{\alpha}^{z}}\overline{\phi_{\alpha ww}})\text{d}^2w\Big],\notag\\
    \langle{T_{z\bar z}}\rangle_1&=-\frac{1}{8\pi G}\Big[\frac{1}{64}e^{2\phi(z,\bar z)}-e^{-2\phi(z,\bar z)}|\partial_z^2\phi-(\partial_z\phi)^2|^2\Big].
\end{align}
A self-consistency check is that $\langle{T_{ij}}\rangle_1$ satisfies both the conservation law (\ref{TTbar deformed conservation equation}) and the $T\bar T$ trace relation (\ref{TTbar deformed trace relation}). Furthermore, since $\partial_{z}^3\Xi_{\alpha}^{z}[\gamma_{\rho}](z)=0$, we can verify that the one-point correlators $\langle{T_{zz}}\rangle_1$ and $\langle{T_{\bar z\bar z}}\rangle_1$ are indeed automorphic forms. The deformed one-point correlators $\langle{T_{zz}}\rangle_1$ and $\langle{T_{\bar z\bar z}}\rangle_1$ involve integrals of the forms $\int e^{-2\phi}\langle{T_{\bar w\bar w}}\rangle_0\partial_z^3(G^z_{ww}+\sum_{\alpha}f_{\alpha}^{z}\phi_{\alpha ww})\text{d}^2w$ and $\int e^{-2\phi}\langle{T_{ww}}\rangle_0\partial_{\bar z}^3(\overline{G^z_{ww}}+\sum_{\alpha}\overline{f_{\alpha}^{z}\phi_{\alpha ww}})\text{d}^2w$, which could suggest the non-locality of the $T\bar T$ deformation.
\subsection{Perturbative stress tensor two-point correlator}\label{subsection Perturbative stress tensor two-point correlator}
According to the generalized GKPW relation (\ref{generalized GKPW relation}), the multi-point stress tensor correlators can be computed by taking functional derivatives of $\langle{T_{ij}}\rangle$ with respect to the boundary metric. We specify the boundary metric as
\begin{align}
    g^{(c)}_{ij}(z,\bar z)=e^{2\phi(z,\bar z)}\eta_{ij}+\epsilon\chi_{ij}(z,\bar z),
\end{align}
where $\epsilon$ is an infinitesimal parameter. The perturbed stress tensor $\langle{T_{ij}(\epsilon)}\rangle_{\rho_c}$ can be written as a {power} series in $\epsilon$, $\langle{T_{ij}(\epsilon)}\rangle_{\rho_c}=\sum_{n=0}^{\infty}\epsilon^n\langle{T_{ij}}\rangle_{\rho_c}^{[n]}$. Expanding the conservation equation (\ref{TTbar deformed conservation equation}) the $T\bar T$ trace relation (\ref{TTbar deformed trace relation}) in $\epsilon$, and the coefficients of $\epsilon^k$ lead to
    \begin{align}
    &\langle{T_{z\bar z}}\rangle_{\rho_c}^{[k]}=\bar A_{\rho_c}\langle{T_{zz}}\rangle_{\rho_c}^{[k]}+A_{\rho_c}\langle{T_{\bar z\bar z}}\rangle_{\rho_c}^{[k]}+\mathcal{F}^{[k]}_{z\bar z\rho_c},\label{trace relation for TTbar}\\
    &\partial_{\bar z}\langle{T_{zz}}\rangle_{\rho_c}^{[k]}=-e^{2\phi}\partial_z(e^{-2\phi}\langle{T_{z\bar z}}\rangle_{\rho_c}^{[k]})+\mathcal{F}^{[k]}_{zz\rho_c},\label{conservation equation 1 for TTbar}\\
    &\partial_{z}\langle{T_{\bar z\bar z}}\rangle_{\rho_c}^{[k]}=-e^{2\phi}\partial_{\bar z}(e^{-2\phi}\langle{T_{z\bar z}}\rangle_{\rho_c}^{[k]})+\mathcal{F}^{[k]}_{\bar z\bar z\rho_c}.\label{conservation equation 2 for TTbar}
\end{align}
Here $\mathcal{F}^{[k]}_{z\bar z\rho_c}$, $\mathcal{F}^{[k]}_{zz\rho_c}$ and $\mathcal{F}^{[k]}_{\bar z\bar z\rho_c}$ consist of lower-order coefficients and local functions of $\chi_{ij}$. $A_{\rho_c}$ and $\bar A_{\rho_c}$ are defined by
\begin{align}
     A_{\rho_c}(z,\bar z)&=\frac{8\pi G\rho_c\langle{T_{zz}(z,\bar z)}\rangle_{\rho_c}^{[0]}}{e^{2\phi(z,\bar z)}+16\pi G\rho_c\langle{T_{z\bar z}(z,\bar z)}\rangle_{\rho_c}^{[0]}},\notag\\
     \bar A_{\rho_c}(z,\bar z)&=\frac{8\pi G\rho_c\langle{T_{\bar z\bar z}(z,\bar z)}\rangle_{\rho_c}^{[0]}}{e^{2\phi(z,\bar z)}+16\pi G\rho_c\langle{T_{z\bar z}(z,\bar z)}\rangle_{\rho_c}^{[0]}}.
\end{align}
Once again, each coefficient $\langle{T_{ij}}\rangle_{\rho_c}^{[n]}$ can be expanded in terms of $\rho_c$. In the leading order, (\ref{trace relation for TTbar})(\ref{conservation equation 1 for TTbar})(\ref{conservation equation 2 for TTbar}) are consistent with the differential equations of CFT correlators. At the subleading order, we obtain
\begin{align}
        \langle{T_{z\bar z}}\rangle_1^{[k]}&=8\pi Ge^{-2\phi}(\langle{T_{\bar z\bar z}}\rangle_0^{[0]}\langle{T_{zz}}\rangle_0^{[k]}+\langle{T_{zz}}\rangle_0^{[0]}\langle{T_{\bar z\bar z}}\rangle_0^{[k]})+\mathcal{F}^{[k]}_{z\bar z1},\\
    \partial_{\bar z}\langle{T_{zz}}\rangle_1^{[k]}&=-8\pi Ge^{2\phi}\partial_z[e^{-4\phi}(\langle{T_{\bar z\bar z}}\rangle_0^{[0]}\langle{T_{zz}}\rangle_0^{[k]}+\langle{T_{zz}}\rangle_0^{[0]}\langle{T_{\bar z\bar z}}\rangle_0^{[k]})]\notag\\
    &\quad+\mathcal{F}^{[k]}_{zz1}-e^{2\phi}\partial_z[e^{-2\phi}\mathcal{F}^{[k]}_{z\bar z1}],\\
     \partial_{z}\langle{T_{\bar z\bar z}}\rangle_1^{[k]}&=-8\pi Ge^{2\phi}\partial_{\bar z}[e^{-4\phi}(\langle{T_{\bar z\bar z}}\rangle_0^{[0]}\langle{T_{zz}}\rangle_0^{[k]}+\langle{T_{zz}}\rangle_0^{[0]}\langle{T_{\bar z\bar z}}\rangle_0^{[k]})]\notag\\
    &\quad+\mathcal{F}^{[k]}_{\bar z\bar z1}-e^{2\phi}\partial_{\bar z}[e^{-2\phi}\mathcal{F}^{[k]}_{z\bar z1}].
\end{align}
One can observe that $\langle{T_{z\bar z}}\rangle_1^{[k]}$, $\langle{T_{zz}}\rangle_1^{[k]}$, and $\langle{T_{\bar z\bar z}}\rangle_1^{[k]}$ are decoupled in these differential equations, which can be solved by employing the genus-$g$ Green's function defined in appendix \ref{appendix Differentials and Green's function}. In this paper,
we present detailed results for the $T\bar{T}$-deformed two-point correlators at the subleading order in the expansion parameter $\rho_c$
\begin{align}
        \langle{T_{z\bar z}T_{ww}}\rangle_1=&\ e^{-2\phi(z,\bar z)}\Big[(\partial_{\bar z}^2\phi-(\partial_{\bar z}\phi)^2)\langle{T_{zz}T_{ww}}\rangle_0+(\partial_{z}^2\phi-(\partial_{z}\phi)^2)\langle{T_{\bar z\bar z}T_{ww}}\rangle_0\Big]\notag\\
    &+\Big[\frac{1}{16\pi^2G}\int_{\mathcal{D}}e^{-2\phi(z',\bar z')}(\partial_{\bar z'}^2\phi-(\partial_{\bar z'}\phi)^2)\partial_z^3(G^{z}_{z'z'}+\sum_{\alpha=1}^{3g-3}f_{\alpha}^{z}\phi_{\alpha z'z'})\text{d}^2z'\notag\\
    &+\frac{1}{64\pi G}(\partial_z^2+2\partial_z\phi\partial_z+5\partial_{z}^2\phi-3(\partial_{z}\phi)^2)\Big]\delta^{(2)}(z-w),\notag\\
         \langle{T_{z\bar z}T_{\bar w\bar w}}\rangle_1=&\text{ c.c. of }\langle{T_{z\bar z}T_{ww}}\rangle_1,\notag\\
    \langle{T_{z\bar z}T_{w\bar w}}\rangle_1=&\ e^{-2\phi(z,\bar z)}\Big[(\partial_{\bar z}^2\phi-(\partial_{\bar z}\phi)^2)\langle{T_{zz}T_{w\bar w}}\rangle_0+(\partial_{z}^2\phi-(\partial_{z}\phi)^2)\langle{T_{\bar z\bar z}T_{w\bar w}}\rangle_0\Big]\notag\\
    &-\frac{1}{32\pi G}\Big[\partial_z\partial_{\bar z}+12e^{-2\phi(z,\bar z)}|\partial_z\phi-(\partial_z\phi)^2|^2-\frac{1}{16}e^{2\phi(z,\bar z)})\Big]\delta^{(2)}(z-w),\notag\\
    \langle{T_{zz}T_{ww}}\rangle_1=&\sum_{\alpha=1}^{3g-3}\phi_{\alpha zz}\frac{\partial}{\partial\tau_{\alpha}}\langle{T_{ww}}\rangle_1-\frac{1}{\pi}\int_{\mathcal{D}}\Big[e^{-2\phi(z',\bar z')}(\partial_{z'}+2\partial_{z'}\phi)G^{z'}_{zz}\notag\\
    &\times[(\partial_{\bar z'}^2\phi-(\partial_{\bar z'}\phi)^2)\langle{T_{z'z'}T_{ww}}\rangle_0+(\partial_{z'}^2\phi-(\partial_{z'}\phi)^2)\langle{T_{\bar z'\bar z'}T_{ww}}\rangle_0]\Big]\text{d}^2z'\notag\\
    &-\frac{1}{16\pi^2G}\Big[\Big(\frac{1}{4}\partial_w^3+(\partial_w^2\phi-(\partial_w\phi)^2)\partial_w+\frac{1}{4}(\partial_w^3\phi+2\partial_w\phi\partial_w^2\phi-4(\partial_w\phi)^3)\Big)G^{w}_{zz}\notag\\
    &-\frac{1}{\pi}\int_{\mathcal{D}}e^{-2\phi(z',\bar z')}(\partial_{z'}^2\phi-(\partial_{z'}\phi)^2)(G^{w}_{zz}\partial_w+2\partial_wG^{w}_{zz})\partial_w^3(G^w_{z'z'}+\sum_{\alpha=1}^{3g-3}f^{w}_{\alpha}\phi_{\alpha z'z'})\text{d}^2z'\Big],\notag\\
    \langle{T_{\bar z\bar z}T_{\bar w\bar w}}\rangle_1=&\text{ c.c. of }\langle{T_{zz}T_{ww}}\rangle_1,\notag\\
    \langle{T_{zz}T_{\bar w\bar w}}\rangle_1=&\sum_{\alpha=1}^{3g-3}\phi_{\alpha zz}\frac{\partial}{\partial\tau_{\alpha}}\langle{T_{\bar w\bar w}}\rangle_1-\frac{1}{\pi}\int_{\mathcal{D}}\Big[e^{-2\phi(z',\bar z')}(\partial_{z'}+2\partial_{z'}\phi)G^{z'}_{zz}\notag\\
    &\times[(\partial_{\bar z'}^2\phi-(\partial_{\bar z'}\phi)^2)\langle{T_{z'z'}T_{\bar w\bar w}}\rangle_0+(\partial_{z'}^2\phi-(\partial_{z'}\phi)^2)\langle{T_{\bar z'\bar z'}T_{\bar w\bar w}}\rangle_0]\Big]\text{d}^2z'\notag\\
    &-\frac{1}{16\pi^2G}\Big[\frac{1}{2}(\partial_{\bar w}^2\phi-(\partial_{\bar w}\phi)^2)(\partial_w+2\partial_w\phi)-\partial^3_{\bar w}[e^{-2\phi(w,\bar w)}(\partial_w^2\phi-(\partial_w\phi)^2)]\Big]G^{w}_{zz}\notag\\
    &-\frac{1}{16\pi G}\Big[\frac{1}{4}(\partial_w\partial_{\bar w}-2\partial_{\bar w}\phi\partial_w+2\partial_{w}\phi\partial_{\bar w})+(\frac{3}{16}e^{2\phi(w,\bar w)}-\partial_w\phi\partial_{\bar w}\phi\notag\\
    &-4e^{-2\phi(w,\bar w)}|\partial_w^2\phi-(\partial_w\phi)^2|^2)\Big](\delta^{(2)}(w-z)-\sum_{\alpha=1}^{3g-3}\mu_{\alpha\bar w}^{w}\phi_{\alpha zz}),
\end{align}
where $\langle{T_{ij}T_{kl}}\rangle_0$ represent the CFT two-point correlator, which has been obtained in subsection \ref{subsection CFT two-point correlators}. Some integral terms that imply the non-locality of the $T\bar T$ deformation can also be found in the two-point stress tensor correlators.
\section{Conclusions and perspectives}\label{section Conclusions and perspectives}
In this paper, we investigate the holographic correlators of stress tensor on a higher genus Riemann surface within the frameworks of AdS$_3$/CFT$_2$ and cutoff-AdS$_3$/$T\bar T$-CFT$_2$, respectively. In AdS$_3$/CFT$_2$, we employ the near-boundary analysis to solve Einstein's equation and utilize the GKPW relation in the semiclassical limit for calculating holographic correlators. We obtain the concrete form of the correlator with up to four stress tensors inserted. In addition, we derive recurrence relations for a specific class of higher-point correlators to establish connections between the $n$-point and $(n+1)$-point correlators. Our results are consistent with the Ward identity in CFT, thus providing a specific validation of AdS$_3$/CFT$_2$ with non-trivial topology. In the context of cutoff-AdS$_3$/$T\bar T$-CFT$_2$, we extend the method of dynamical coordinates to the Riemann surface. We provide a construction of dynamical coordinate transformation that ensures the single-valuedness of deformed stress tensor correlators on the Riemann surface. Subsequently, we employ the perturbation method to calculate the deformed one-point and two-point stress tensor correlators at the subleading order in the deformation parameter.\par
The results in this paper apply to the case where the Euclidean space is a handlebody, and it would be interesting to extend our calculations to non-handlebody solutions, such as the solution described in \cite{Maldacena:2004rf}. Furthermore, investigating holographic correlators in the presence of matter fields in the bulk is also a crucial direction. Additionally, it is imperative to develop a non-perturbative approach for computing holographic correlators in cutoff-AdS$_3$. It is also possible to calculate the correlators of KdV charges on higher genus Riemann surfaces. Generalized KdV charges on such surfaces have already been considered in \cite{Bonora:1988wn}.\par
It should be mentioned again that during the calculation we assume that only one saddle dominates, which holds true at a certain limit. In order to get a better understanding of $\mathrm{AdS}_3/\mathrm{CFT}_2$ in the semiclassical limit, it is crucial to find a way to take account of contributions to holographic correlators from the complete summation over saddle points. Recently, a new contribution beyond the $\mathrm{SL}(2,\mathbb{Z})$ saddle summation has been introduced to construct a unitary torus partition function \cite{DiUbaldo:2023hkc}, which corresponds to a spinning string coupled to gravity in $\mathrm{AdS}_3$. It is also tempting to explore how this new configuration will influence the behavior of holographic correlators in the boundary theory, and how the picture can be extended to the higher genus case.

\section*{Acknowledgments}
We want to thank Yi Li, and Yunda Zhang for useful discussions related to this work. S.H. also would like to appreciate the financial support from the Max Planck Partner Group, the Fundamental Research Funds for the Central Universities, and the Natural Science Foundation of China Grants No.~12075101, No.~12235016.

\appendix

\section{Differentials and Green's function}\label{appendix Differentials and Green's function}
We start by reviewing certain aspects of differentials on a Riemann surface. After introducing the metric $\text{d}s^2=g_{ab}\text{d}\xi^a\text{d}\xi^b$ on the Riemann surface, the compatible complex structure is defined as follows:
\begin{align}
    J_a^b=\sqrt{g}\,\varepsilon_{ac}g^{cb},
\end{align}
where $\varepsilon_{11}=\varepsilon_{22}=0$, $\varepsilon_{12}=-\varepsilon_{21}=1$. Subsequently, one can establish harmonic coordinates $(z,\bar z)$ that satisfy the Beltrami equation,
\begin{align}
    J^b_{a}\frac{\partial z}{\partial\xi^b}=i\frac{\partial z}{\partial\xi^a},\ \ \ J^b_{a}\frac{\partial \bar z}{\partial\xi^b}=-i\frac{\partial \bar z}{\partial\xi^a},
\end{align}
in which the metric takes the form
\begin{align}
    \text{d}s^2=\rho(z,\bar z)\text{d}z\text{d}\bar z.
\end{align}
The space of the metrics on a genus-$g$ Riemann surface is denoted as $\mathscr{G}_g$. The variations of the metric can be classified into two categories, with the first category being unphysical and encompassing diffeomorphisms and Weyl transformations,
\begin{align}
    \delta\tilde{g}_{ij}\text{d}z^i\text{d}z^j=\rho\delta\tilde\omega\text{d}z\text{d}\bar z+\rho\partial_{z}\epsilon^{\bar z}(\text{d}z)^2+\rho\partial_{\bar z}\epsilon^{z}(\text{d}\bar z)^2,
\end{align}
where $\delta\tilde\omega=\delta\omega+\partial_{z}(\rho\epsilon^z)+\partial_{\bar z}(\rho\epsilon^{\bar z})$ for some Weyl rescaling $\delta\omega$ and infinitesimal vector field $\epsilon^i$. Typically, selecting a gauge slice $\Sigma$ in $\mathscr{G}_g$ is necessary to fix the unphysical degrees of freedom, and the variations tangent to the gauge slice are considered as physically meaningful.
The physical variation is denoted as $\delta g_{ij}$, and it can be formally written as 
\begin{align}
    \delta{g}_{ij}\text{d}z^i\text{d}z^j=\rho\delta\varphi\text{d}z\text{d}\bar z+\delta\phi_{zz}(\text{d}z)^2+\overline{\delta\phi_{zz}}(\text{d}\bar z)^2.
\end{align}
Applying the orthogonality condition introduced in \cite{Belavin:1986ga},
\begin{align}
    \|\delta\tilde{g},\delta{g}\|&=\int\sqrt{g}g^{ik}g^{jl}\delta\tilde g_{ij}\delta g_{kl}\text{d}^2z\notag\\
    &=\int\Big(\rho\delta\tilde\omega\delta\varphi+\delta\phi_{zz}\partial_{\bar z}\epsilon^{z}+\overline{\delta\phi_{zz}}\partial_{z}\epsilon^{\bar z}\Big)\text{d}^2z\notag\\
    &=0,
\end{align}
we obtain
\begin{align}
    \delta\varphi=0,\ \ \ \partial_{\bar z}\delta\phi_{zz}=\partial_z\overline{\delta\phi_{zz}}=0.
\end{align}
The physical variation $\delta \phi_{zz}(z)(dz)^2$ is known as the holomorphic quadratic differential. According to the Riemann-Roch theorem \cite{Nakahara:2003nw}, the dimension of the linear space $\mathcal{H}^{2}_{g}$ for holomorphic quadratic differentials is $3g-3$ (when $g\geq 2$, and $1$ when $g=1$). Thus $\delta\phi_{zz}(z)$ can be parameterized by the variations of $3g-3$ complex modular parameters $\lbrace{\bar\tau_\alpha}\rbrace$,
\begin{align}
    \delta\phi_{zz}(z)=\sum_{\alpha=1}^{3g-3}\phi_{\alpha zz}(z)\delta\bar\tau_\alpha.
\end{align}
After selecting a basis $\lbrace{\phi_{\alpha zz}}\rbrace$ in $\mathcal{H}^{2}_{g}$, the dual basis in the space of Beltrami differentials is defined as follows:
\begin{align}
    \int\phi_{\alpha zz}(z)\mu^z_{\beta\bar z}(z,\bar z)\text{d}^2z=\delta_{\alpha\beta},
\end{align}
where $\delta_{\alpha\beta}=1$ for $\alpha=\beta$ and $\delta_{\alpha\beta}=0$ for $\alpha\neq\beta$. The construction of the basis $\lbrace{\mu^z_{\alpha\bar z}}\rbrace$ is provided in \cite{Belavin:1986ga},
\begin{align}
    \mu^z_{\alpha\bar z}=&\sum_{\beta=1}^{3g-3}\rho^{-1}(N_2^{-1})_{\alpha\beta}\overline{\phi_{\beta zz}},\notag\\
    \text{where}\ \ \ (N_2)_{\alpha\beta}=&\int\rho^{-1}(z,\bar z)\overline{\phi_{\alpha zz}}(\bar z)\phi_{\beta zz}(z)\text{d}^2z. \label{dual basis of Beltrami differentials}
\end{align}
By choosing an appropriate basis\footnote{Given any basis for the space of holomorphic quadratic differentials, we can employ the Gram-Schmidt orthogonalization to find a new basis satisfies $(N_2)_{\alpha\beta}=\delta_{\alpha\beta}$.} $\lbrace{\phi_{\alpha zz}}\rbrace$ such that $(N_2)_{\alpha\beta}=\delta_{\alpha\beta}$, the dual Beltrami differential can be simplified as $\mu^z_{\alpha\bar z}=\rho^{-1}\overline{\phi_{\alpha zz}}$, as employed in \cite{Martinec:1986bq,DHoker:1988pdl,Giddings:1987im,Roland:1993pm}. The Beltrami differentials naturally parameterize the metrics on a Riemann surface. Select a point on the gauge slice $\Sigma$ equipped with the metric $\text{d}s^2(\tau)=\rho(\tau)\text{d}z\text{d}\bar z$, and the metric at the neighboring point $\tau+\delta\tau$ can be expressed as
\begin{align}
    \text{d}s^2(\tau+\delta\tau)=\rho(\tau)\Big|\text{d}z+\sum_{\alpha=1}^{3g-3}\mu^z_{\alpha\bar z}(\tau)\delta\tau_\alpha\text{d}\bar z\Big|^2.
\end{align}\par 
On a genus-$g$ Riemann surface, the Green's function $G_{N}(z,\bar z;w,\bar w)$ for $\partial_{\bar z}$ is a bidifferential of weight $(1-N,N)$ satisfies the following two equations \cite{Tuite:2019fqx},
\begin{align}
    \frac{1}{\pi}\partial_{\bar z}G^{\scaleto{\overbrace{z...z}^{N-1}}{13pt}}_{\scaleto{\underbrace{w...w}_N}{13pt}}(z,\bar z;w,\bar w)&=\delta(z-w)-p_{N}(z,\bar z;w),\notag\\
     \frac{1}{\pi}\partial_{\bar w}G^{\scaleto{\overbrace{z...z}^{N-1}}{13pt}}_{\scaleto{\underbrace{w...w}_N}{13pt}}(z,\bar z;w,\bar w)&=-\delta(z-w). \label{Green's function's definition}
\end{align}
Here $p_{N}(z,w)$ is the projection kernel defined as
\begin{align}
    p_{N}(z,\bar z;w)=&\sum_{\alpha=1}^{\text{dim}\mathcal{H}_g^N}{\phi}^*_{\alpha\scaleto{\underbrace{w...w}_{N}}{13pt}}(w)\overline{\phi_{\alpha\scaleto{\underbrace{z...z}_{N}}{13pt}}}(\bar z)\rho^{1-N}(z,\bar z),
\end{align}
where $\lbrace{\phi_{\alpha\scaleto{\underbrace{z...z}_{N}}{13pt}}}\rbrace$ is a basis of the space $\mathcal{H}_g^N$ for holomorphic $N$-differentials and $\lbrace{\phi^*_{\alpha \scaleto{\underbrace{w...w}_{N}}{13pt}}}\rbrace$ is the dual basis of $\lbrace{\phi_{\alpha \scaleto{\underbrace{w...w}_{N}}{13pt}}}\rbrace$ with respect to the Petersson inner product,
\begin{align}
    \langle{{\phi}^*_{\alpha\scaleto{\underbrace{w...w}_{N}}{13pt}},{\phi}_{\beta\scaleto{\underbrace{w...w}_{N}}{13pt}}}\rangle=&\int\rho^{1-N}(w,\bar w)\phi^*_{\alpha\scaleto{\underbrace{w...w}_{N}}{13pt}}(w)\overline{\phi_{\beta\scaleto{\underbrace{w...w}_{N}}{13pt}}}(\bar w)\text{d}^2w=\delta_{\alpha\beta}.
\end{align}
In this paper, we are concerned with the case of $N=2$. Since we have already assumed that $(N_2)_{\alpha\beta}=\delta_{\alpha\beta}$, the Petersson dual $\phi_{\alpha ww}^*$ is equivalent to $\phi_{\alpha ww}$. The complex conjugate $\overline{\phi_{\alpha zz}}$ is further substituted with the Beltrami differential $\mu_{\alpha\bar z}^z$, yielding \cite{Eguchi:1986sb,Tuite:2019fqx}
\begin{align}
    p_2(z,\bar z;w)&=\sum_{\alpha=1}^{3g-3}\mu_{\alpha\bar z}^z(z,\bar z)\phi_{\alpha ww}(w).
\end{align}
For a general basis $\lbrace{\phi_{\alpha zz}}\rbrace$ the projection kernel takes the form
\begin{align}
    p_2(z,\bar z;w)=\sum_{\alpha,\beta=1}^{3g-3}[(N_2)_{\alpha\beta}\mu_{\beta\bar z}^{z}](z,\bar z)\phi^*_{\alpha ww}(w).
\end{align}
When employing Schottky uniformization, the exact Green's function can be expressed by utilizing the Poincare series in the following manner \cite{McIntyre:2004xs,Tuite:2019fqx,Martinec:1986bq,DiVecchia:1989id}:
\begin{align}
    G^{\scaleto{\overbrace{z...z}^{N-1}}{13pt}}_{\scaleto{\underbrace{w...w}_N}{13pt}}(z,\bar z;w,\bar w)=&-\sum_{\gamma\in\Gamma}(\gamma'(w))^N\frac{1}{\gamma(w)-z}\prod_{j=1}^{2N-1}\frac{z-A_j}{\gamma(w)-A_j}\notag\\
    &-\sum_{\alpha=1}^{\text{dim}\mathcal{H}^N_g}\phi^*_{\alpha\scaleto{\underbrace{w...w}_{N}}{13pt}}(w)f^{\scaleto{\overbrace{z...z}^{N-1}}{13pt}}_{\alpha}(z),\label{PS definition of Green's Function}
\end{align}
where $\Gamma$ is the Schottky group and $\lbrace{A_j}\rbrace$ are distinct elements of the limit set $\Lambda(\Gamma)$. $f^{\scaleto{\overbrace{z...z}^{N-1}}{13pt}}_{\alpha}$ is the Bers potential for $\phi_{\alpha\scaleto{\underbrace{z...z}_{N}}{13pt}}$ \cite{Bers1967,Bers1971}, which can be constructed as
\begin{align}
    f^{\scaleto{\overbrace{z...z}^{N-1}}{13pt}}_{\alpha}=-\int\rho^{1-N}(w,\bar w)\sum_{\gamma\in\Gamma}(\gamma'(w))^N\frac{1}{\gamma(w)-z}\prod_{j=1}^{2N-1}\frac{z-A_j}{\gamma(w)-A_j}\overline{\phi_{\alpha\scaleto{\underbrace{w...w}_{N}}{13pt}}}\text{d}^2w.
\end{align}
The second term on the right-hand side of (\ref{PS definition of Green's Function}) ensures that $G^{\scaleto{\overbrace{z...z}^{N-1}}{13pt}}_{\scaleto{\underbrace{w...w}_N}{13pt}}$ is an automorphic form in both $z$ and $w$, i.e. satisfies \cite{McIntyre:2004xs}
\begin{align}
    G^{\scaleto{\overbrace{z...z}^{N-1}}{13pt}}_{\scaleto{\underbrace{w...w}_N}{13pt}}(\gamma(z),\overline{\gamma(z)};w,\bar w)(\gamma'(z))^{1-N}&=G^{\scaleto{\overbrace{z...z}^{N-1}}{13pt}}_{\scaleto{\underbrace{w...w}_N}{13pt}}(z,\bar z;w,\bar w),\notag\\
    G^{\scaleto{\overbrace{z...z}^{N-1}}{13pt}}_{\scaleto{\underbrace{w...w}_N}{13pt}}(z,\bar z;\gamma(w),\overline{\gamma(w)})(\gamma'(w))^{N}&=G^{\scaleto{\overbrace{z...z}^{N-1}}{13pt}}_{\scaleto{\underbrace{w...w}_N}{13pt}}(z,\bar z;w,\bar w).
\end{align}

\section{List of three-point and four-point correlators}\label{appendix List of three-point and four-point correlators}
We show the list of all six independent three-point correlators and nine independent four-point correlators of the CFT case in this appendix. Other correlators at the same order can be obtained by complex conjugation. For simplicity, we will use $\delta_i$ and $\delta_{ij}$ as the abbreviation of $\delta^{(2)}(z-z_i)$ and $\delta^{(2)}(z_i-z_j)$. The six three-point correlators are:
\begin{align}
    &\braket{T_{zz}(z)T_{zz}(z_1)T_{zz}(z_2)}=\frac{2}{\pi}\sum\limits_{i=1}^2\partial_{z_i}G^{z_i}_{\ zz}(z_i,\Bar{z}_i;z,\Bar{z})\braket{T_{zz}(z_1)T_{zz}(z_2)}\nonumber\\
    &\quad\quad+\frac{1}{\pi}\sum\limits_{i=1}^2 G^{z_i}_{\ zz}(z_i,\Bar{z}_i;z,\Bar{z})\partial_{z_i}\braket{T_{zz}(z_1)T_{zz}(z_2)}+\sum\limits_{\alpha=1}^{3g-3}\phi_{\alpha zz}(z)\frac{\partial}{\partial \tau_\alpha}\braket{T_{zz}(z_1)T_{zz}(z_2)},\\
    &\braket{T_{zz}(z)T_{\Bar{z}\Bar{z}}(z_1)T_{\Bar{z}\Bar{z}}(z_2)}=\frac{1}{\pi}\sum\limits_{i=1}^2G^{z_i}_{\ zz}(z_i,\Bar{z}_i;z,\Bar{z})\partial_{z_i}\braket{T_{\Bar{z}\Bar{z}}(z_1)T_{\Bar{z}\Bar{z}}(z_2)}\nonumber\\
    &\ +\frac{1}{16\pi^2 G}\sum\limits_{i=1}^2\Big[G^{z_i}_{\ zz}(z_i,\Bar{z}_i;z,\Bar{z})\big(-8\partial_{\Bar{z}_i}^2\phi\partial_{\Bar{z}_i}+8\partial_{\Bar{z}_i}\phi\partial_{\Bar{z}_i}^2-4\partial_{\Bar{z}_i}^3\phi+24\partial_{\Bar{z_i}}\phi\partial_{\Bar{z}_i}^2\phi-16(\partial_{\Bar{z}_i}\phi)^3\big)\nonumber\\
    &\ +2\partial_{\Bar{z}_i}G^{z_i}_{\ zz}(z_i,\Bar{z}_i;z,\Bar{z})\big(\partial_{\Bar{z}_i}^2-4\partial_{\Bar{z}_i}^2\phi\partial_{\Bar{z}_i}\big)\Big]\delta_{12}+\sum\limits_{\alpha=1}^{3g-3}\phi_{\alpha zz}\frac{\partial}{\partial \tau_\alpha}\braket{T_{\Bar{z}\Bar{z}}(z_1)T_{\Bar{z}\Bar{z}}(z_2)},\\
    &\braket{T_{z\Bar{z}}(z)T_{zz}(z_1)T_{zz}(z_2)}=\braket{T_{zz}(z)T_{zz}(z_2)}\delta_1+\braket{T_{zz}(z)T_{zz}(z_1)}\delta_2,
\end{align}
\begin{align}
    &\braket{T_{z\Bar{z}}(z)T_{zz}(z_1)T_{\Bar{z}\Bar{z}}(z_2)}=\braket{T_{zz}(z)T_{\Bar{z}\Bar{z}}(z_2)}\delta_1+\braket{T_{\Bar{z}\Bar{z}}(z)T_{zz}(z_1)}\delta_2\nonumber\\
    &\quad\quad\quad+\frac{1}{16\pi G}(\partial_z\delta_2\partial_{\Bar{z}}\delta_1-\partial_z\delta_1\partial_{\Bar{z}}\delta_2-4\delta_2\partial_{\Bar{z}}\delta_1\partial_z\phi-4\delta_1\partial_z\delta_2\partial_{\Bar{z}}\phi+16\delta_1\delta_2\partial_z\phi\partial_{\Bar{z}}\phi),\\[2mm]
    &\braket{T_{z\Bar{z}}(z)T_{z\Bar{z}}(z_1)T_{zz}(z_2)}=\braket{T_{z\Bar{z}}(z)T_{zz}(z_2)}\delta_1+\braket{T_{zz}(z)T_{z\Bar{z}}(z_1)}\delta_2\nonumber\\
    &\quad+\frac{1}{16\pi G}\big(4(\partial_z\phi)^2\delta_1\delta_2+4\delta_1\partial_z\delta_2\partial_z\phi-\partial_z\delta_1\partial_z\delta_2-4\delta_1\delta_2\partial_z^2\phi-\delta_1\partial_z^2\delta_2\big),\\[2mm]
    &\braket{T_{z\Bar{z}}(z)T_{z\Bar{z}}(z_1)T_{z\Bar{z}}(z_2)}=2\braket{T_{z\Bar{z}}(z)T_{z\Bar{z}}(z_1)}\delta_2+2\braket{T_{z\Bar{z}}(z)T_{z\Bar{z}}(z_2)}\delta_1\nonumber\\
    &\ +\frac{1}{16\pi G}\Big[24\delta_1\delta_2\partial_z\phi\partial_{\Bar{z}}\phi-6\delta_1\big(\partial_z\phi\partial_{\Bar{z}}\delta_2+\partial_{\Bar{z}}\phi\partial_z\delta_2\big)-6\delta_2\big(\partial_z\phi\partial_{\Bar{z}}\delta_1+\partial_{\Bar{z}}\phi\partial_z\delta_1\big)\nonumber\\
    &\quad+\partial_z\delta_1\partial_{\Bar{z}}\delta_2+\partial_z\delta_2\partial_{\Bar{z}}\delta_1-4\delta_1\delta_2\partial_z\partial_{\Bar{z}}\phi+2\delta_1\partial_z\partial_{\Bar{z}}\delta_2++2\delta_2\partial_z\partial_{\Bar{z}}\delta_1\Big].
\end{align}
And the nine four-point correlators are:
\begin{align}
    &\braket{T_{zz}(z)T_{zz}(z_1)T_{zz}(z_2)T_{zz}(z_3)}=\frac{2}{\pi}\sum\limits_{i=1}^3\partial_{z_i}G^{z_i}_{\ zz}(z_i,\Bar{z}_i;z,\Bar{z})\braket{T_{zz}(z_1)T_{zz}(z_2)T_{zz}(z_3)}\nonumber\\
    &+\frac{1}{\pi}\sum\limits_{i=1}^3 G^{z_i}_{\ zz}(z_i,\Bar{z}_i;z,\Bar{z})\partial_{z_i}\braket{T_{zz}(z_1)T_{zz}(z_2)T_{zz}(z_3)}+\sum\limits_{\alpha=1}^{3g-3}\phi_{\alpha zz}(z)\frac{\partial}{\partial \tau_\alpha}\braket{T_{zz}(z_1)T_{zz}(z_2)T_{zz}(z_3)},\\
    &\braket{T_{zz}(z)T_{\Bar{z}\Bar{z}}(z_1)T_{\Bar{z}\Bar{z}}(z_2)T_{\Bar{z}\Bar{z}}(z_3)}=\frac{1}{\pi}\sum\limits_{i=1}^3G^{z_i}_{\ zz}(z_i,\Bar{z}_i;z,\Bar{z})\partial_{\Bar{z}_i}\braket{T_{\Bar{z}\Bar{z}}(z_1)T_{\Bar{z}\Bar{z}}(z_2)T_{\Bar{z}\Bar{z}}(z_3)}\nonumber\\
    &-\frac{2}{\pi}\sum\limits_{\sigma\in S_3}G^{z_{\sigma(1)}}_{\ zz}(z_{\sigma(1)},\Bar{z}_{\sigma(1)};z,\Bar{z})\Big(2\braket{T_{\Bar{z}\Bar{z}}(z_{\sigma(1)})T_{\Bar{z}\Bar{z}}(z_{\sigma(2)})}\partial_{\Bar{z}_{\sigma(1)}}+\partial_{\Bar{z}_{\sigma(1)}}\braket{T_{\Bar{z}\Bar{z}}(z_{\sigma(1)})T_{\Bar{z}\Bar{z}}(z_{\sigma(2)})}\nonumber\\
    &+4\braket{T_{\Bar{z}\Bar{z}}(z_{\sigma(1)})T_{\Bar{z}\Bar{z}}(z_{\sigma(2)})}\partial_{\Bar{z}_{\sigma(1)}}\phi\Big)\delta_{\sigma(1)\sigma(3)}+\sum\limits_{\alpha=1}^{3g-3}\phi_{\alpha zz}\frac{\partial}{\partial \tau_\alpha}\braket{T_{\Bar{z}\Bar{z}}(z_1)T_{\Bar{z}\Bar{z}}(z_2)T_{\Bar{z}\Bar{z}}(z_3)},\\[2mm]
    &\braket{T_{z\Bar{z}}(z)T_{zz}(z_1)T_{zz}(z_2)T_{zz}(z_3)}=\braket{T_{zz}(z)T_{zz}(z_2)T_{zz}(z_3)}\delta^{(2)}(z-z_1)\nonumber\\
    &+\braket{T_{zz}(z)T_{zz}(z_1)T_{zz}(z_3)}\delta^{(2)}(z-z_2)+\braket{T_{zz}(z)T_{zz}(z_1)T_{zz}(z_2)}\delta^{(2)}(z-z_3),\\[2mm]
    &\braket{T_{z\Bar{z}}(z)T_{z\Bar{z}}(z_1)T_{z\Bar{z}}(z_2)T_{zz}(z_3)}=\braket{T_{zz}(z)T_{z\Bar{z}}(z_1)T_{z\Bar{z}}(z_2)}\delta_3+\braket{T_{z\Bar{z}}(z)T_{z\Bar{z}}(z_2)T_{zz}(z_3)}\delta_1\nonumber\\
    &+\braket{T_{z\Bar{z}}(z)T_{z\Bar{z}}(z_1)T_{zz}(z_3)}\delta_2-2\braket{T_{zz}(z)T_{z\Bar{z}}(z_2)}\delta_1\delta_3-2\braket{T_{zz}(z)T_{z\Bar{z}}(z_1)}\delta_2\delta_3\nonumber\\
    &+\frac{1}{16\pi G}\Big[-4\delta_1\delta_2\delta_3(\partial_z\phi)^2-12\delta_1\delta_2\partial_z\phi\partial_z\delta_3+2\delta_1\partial_z\delta_2\partial_z\delta_3+2\delta_2\partial_z\delta_1\partial_z\delta_3+4\delta_1\delta_2\delta_3\partial_z^2\phi\nonumber\\
    &+2\delta_1\delta_2\partial_z^2\delta_3\Big],\\[2mm]
    &\braket{T_{z\Bar{z}}(z)T_{zz}(z_1)T_{zz}(z_2)T_{\Bar{z}\Bar{z}}(z_3)}=\braket{T_{zz}(z)T_{zz}(z_1)T_{\Bar{z}\Bar{z}}(z_3)}\delta_2+\braket{T_{zz}(z)T_{zz}(z_2)T_{\Bar{z}\Bar{z}}(z_3)}\delta_1\nonumber\\
    &+\braket{T_{\Bar{z}\Bar{z}}(z)T_{zz}(z_1)T_{zz}(z_2)}\delta_3+\frac{1}{16\pi G}\Big[-8\delta_3\delta_1\partial_z\delta_2\partial_z\phi-8\delta_3\delta_2\partial_z\delta_1\partial_z\phi+2\delta_1\partial_z\delta_2\partial_z\delta_3\nonumber\\
    &+2\delta_2\partial_z\delta_1\partial_z\delta_3+4\delta_3\partial_z\delta_1\partial_z\delta_2\Big],\\[2mm]
    &\braket{T_{z\Bar{z}}(z)T_{z\Bar{z}}(z_1)T_{zz}(z_2)T_{zz}(z_3)}=\braket{T_{zz}(z)T_{z\Bar{z}}(z_1)T_{zz}(z_2)}\delta_3+\braket{T_{zz}(z)T_{z\Bar{z}}(z_1)T_{zz}(z_3)}\delta_2\nonumber\\
    &-2\braket{T_{zz}(z)T_{zz}(z_2)}\delta_1\delta_3-2\braket{T_{zz}(z)T_{zz}(z_3)}\delta_1\delta_2,
\end{align}
\begin{align}
    &\braket{T_{z\Bar{z}}(z)T_{z\Bar{z}}(z_1)T_{zz}(z_2)T_{\Bar{z}\Bar{z}}(z_3)}=\braket{T_{zz}(z)T_{z\Bar{z}}(z_1)T_{\Bar{z}\Bar{z}}(z_3)}\delta_2+\braket{T_{\Bar{z}\Bar{z}}(z)T_{z\Bar{z}}(z_1)T_{zz}(z_2)}\delta_3\nonumber\\
    &-\braket{T_{zz}(z)T_{\Bar{z}\Bar{z}}(z_3)}\delta_1\delta_2-\braket{T_{\Bar{z}\Bar{z}}(z)T_{zz}(z_2)}\delta_1\delta_3+\frac{1}{16\pi G}\Big[-48\delta_1\delta_2\delta_3\partial_z\phi\partial_{\Bar{z}}\phi+8\delta_1\delta_3\partial_{\Bar{z}}\delta_2\partial_z\phi\nonumber\\
    &+8\delta_2\delta_3\partial_{\Bar{z}}\delta_1\partial_z\phi+8\delta_1\delta_2\partial_z\delta_3\partial_{\Bar{z}}\phi+8\delta_2\delta_3\partial_z\delta_1\partial_{\Bar{z}}\phi+\delta_1\partial_z\delta_2\partial_{\Bar{z}}\delta_3-\delta_1\partial_{\Bar{z}}\delta_2\partial_z\delta_3-2\delta_2\partial_{\Bar{z}}\delta_1\partial_z\delta_3\nonumber\\
    &-2\delta_3\partial_{\Bar{z}}\delta_2\partial_z\delta_1\Big],\\[2mm]
    &\braket{T_{z\Bar{z}}(z)T_{z\Bar{z}}(z_1)T_{z\Bar{z}}(z_2)T_{z\Bar{z}}(z_3)}=2\braket{T_{z\Bar{z}}(z)T_{z\Bar{z}}(z_2)T_{z\Bar{z}}(z_3)}\delta_1+2\braket{T_{z\Bar{z}}(z)T_{z\Bar{z}}(z_1)T_{z\Bar{z}}(z_3)}\delta_2\nonumber\\
    &+2\braket{T_{z\Bar{z}}(z)T_{z\Bar{z}}(z_1)T_{z\Bar{z}}(z_2)}(\delta_3+\delta_{13}+\delta_{23})-6\braket{T_{z\Bar{z}}(z)T_{z\Bar{z}}(z_1)}\delta_2\delta_3\nonumber\\
    &-6\braket{T_{z\Bar{z}}(z)T_{z\Bar{z}}(z_2)}\delta_1\delta_3-4\braket{T_{z\Bar{z}}(z)T_{z\Bar{z}}(z_3)}\delta_1\delta_2+6\braket{T_{z\Bar{z}}(z)}\delta_1\delta_2\delta_3\nonumber\\
    &+\frac{1}{16\pi G}\sum\limits_{\sigma\in S_3}\Big[-24\delta_{\sigma(1)}\delta_{\sigma(2)}\delta_{\sigma(3)}\partial_z\phi\partial_{\Bar{z}}\phi+12\delta_{\sigma(1)}\delta_{\sigma(2)}\partial_{\Bar{z}}\delta_{\sigma(3)}\partial_z\phi\nonumber\\
    &+12\delta_{\sigma(1)}\delta_{\sigma(2)}\partial_z\delta_{\sigma(3)}\partial_{\Bar{z}}\phi-3\delta_{\sigma(1)}\partial_z\delta_{\sigma(2)}\partial_{\Bar{z}}\delta_{\sigma(3)}+6\delta_{\sigma(1)}\delta_{\sigma(2)}\delta_{\sigma(3)}\partial_z\partial_{\Bar{z}}\phi\nonumber\\
    &-3\delta_{\sigma(1)}\delta_{\sigma(2)}\partial_z\partial_{\Bar{z}}\delta_{\sigma(3)}\Big],\\[2mm]
    &\braket{T_{zz}(z)T_{zz}(z_1)T_{\Bar{z}\Bar{z}}(z_2)T_{\Bar{z}\Bar{z}}(z_3)}=\frac{1}{\pi}\big[-4G^{z_2}_{\ zz}(z_2,\Bar{z}_2;z,\Bar{z})\partial_{\Bar{z}_2}\braket{T_{zz}(z_1)T_{\Bar{z\Bar{z}}}(z_2)}\delta_{23}\nonumber\\
    &+2G^{z_1}_{\ zz}(z_1,\Bar{z}_1;z,\Bar{z})\partial_{\Bar{z}_1}\braket{T_{zz}(z_1)T_{\Bar{z}\Bar{z}}(z_2)}\delta_{13}+2G^{z_1}_{\ zz}(z_1,\Bar{z}_1;z,\Bar{z})\partial_{\Bar{z}_1}\braket{T_{zz}(z_1)T_{\Bar{z}\Bar{z}}(z_3)}\delta_{12}\nonumber\\
    &+16G^{z_2}_{\ zz}(z_2,\Bar{z}_2;z,\Bar{z})\braket{T_{zz}(z_1)T_{\Bar{z}\Bar{z}}(z_2)}\delta_{23}\partial_{\Bar{z}_2}\phi-4G^{z_2}_{\ zz}(z_2,\Bar{z}_2;z,\Bar{z})\braket{T_{zz}(z_1)T_{\Bar{z}\Bar{z}}(z_2)}\partial_{\Bar{z}_2}\delta_{23}\nonumber\\
    &-4G^{z_3}_{\ zz}(z_3,\Bar{z}_3;z,\Bar{z})\braket{T_{zz}(z_1)T_{\Bar{z}\Bar{z}}(z_3)}\partial_{\Bar{z}_3}\delta_{23}+G^{z_3}_{\ zz}(z_3,\Bar{z}_3;z,\Bar{z})\partial_{z_3}\braket{T_{zz}(z_1)T_{\Bar{z}\Bar{z}}(z_2)T_{\Bar{z}\Bar{z}}(z_3)}\nonumber\\
    &+G^{z_2}_{\ zz}(z_2,\Bar{z}_2;z,\Bar{z})\partial_{z_2}\braket{T_{zz}(z_1)T_{\Bar{z}\Bar{z}}(z_2)T_{\Bar{z}\Bar{z}}(z_3)}-G^{z_1}_{\ zz}(z_1,\Bar{z}_1;z,\Bar{z})\partial_{z_1}\braket{T_{zz}(z_1)T_{\Bar{z}\Bar{z}}(z_2)T_{\Bar{z}\Bar{z}}(z_3)}\nonumber\\
    &+4G^{z_1}_{\ zz}(z_1,\Bar{z}_1;z,\Bar{z})\braket{T_{zz}(z_1)T_{\Bar{z}\Bar{z}}(z_2)T_{\Bar{z}\Bar{z}}(z_3)}\partial_{z_1}\phi+2G^{z_1}_{\ zz}(z_1,\Bar{z}_1;z,\Bar{z})\partial_{z_1}\braket{T_{zz}(z_1)T_{\Bar{z}\Bar{z}}(z_2)T_{\Bar{z}\Bar{z}}(z_3)}\nonumber\\
    &+2\partial_{z_1}G^{z_1}_{\ zz}(z_1,\Bar{z}_1;z,\Bar{z})\braket{T_{zz}(z_1)T_{\Bar{z}\Bar{z}}(z_2)T_{\Bar{z}\Bar{z}}(z_3)}\big]\nonumber\\
    &+\frac{1}{16\pi^2 G}\Big\{G^{z_2}_{\ zz}(z_2,\Bar{z}_2;z,\Bar{z})\Big[208\delta_{12}\delta_{23}(\partial_{\Bar{z}_2}\phi)^2\partial_{z_2}\phi-80\delta_{23}\partial_{\Bar{z}_2}\delta_{12}\partial_{z_2}\phi\partial_{\Bar{z}_2}\phi\nonumber\\
    &+80\delta_{12}\partial_{\Bar{z}_2}\delta_{23}\partial_{z_2}\phi\partial_{\Bar{z}_2}\phi+16\partial_{\Bar{z}_2}(\delta_{12}\delta_{23}\partial_{z_2}\phi)-48\delta_{12}\delta_{23}\partial_{z_2}\phi\partial_{\Bar{z}_2}^2\phi+8\delta_{23}\partial_{\Bar{z}_2}^2\delta_{12}\partial_{z_2}\phi\nonumber\\
    &-40\delta_{12}\partial_{\Bar{z}}^2\delta_{23}\partial_{z_2}\phi-4\partial_{\Bar{z}}^2\delta_{23}\partial_{z_2}\delta_{12}-32\partial_{\Bar{z}_2}\delta_{23}\partial_{z_2}\delta_{12}\partial_{\Bar{z}_2}\phi-4\partial_{\Bar{z}_2}(\partial_{z_2}\delta_{12}\partial_{\Bar{z}_2}\delta_{23})\nonumber\\
    &+4\delta_{23}\partial_{z_2}\delta_{12}\partial_{\Bar{z}_2}^2\phi+8\partial_{\Bar{z}_2}^2\delta_{23}\partial_{z_2}\delta_{12}-52\delta_{12}\partial_{z_2}\delta_{23}(\partial_{\Bar{z}_2}\phi)^2+20\partial_{z_2}\delta_{23}\partial_{\Bar{z}_2}\delta_{12}\partial_{\Bar{z}_2}\phi\nonumber\\
    &+12\partial_{z_2}(\delta_{12}\partial_{\Bar{z}_2}\delta_{23}\partial_{\Bar{z}_2}\phi)+12\delta_{12}\partial_{z_2}\delta_{23}\partial_{\Bar{z}_2}^2\phi-2\partial_{z_2}\delta_{23}\partial_{\Bar{z}_2}^2\delta_{12}-6\partial_{z_2}(\delta_{12}\partial_{\Bar{z}_2}^2\delta_{23})\nonumber\\
    &-112\delta_{12}\delta_{23}\partial_{\Bar{z}_2}\phi\partial_{z_2}\partial_{\Bar{z}_2}\phi+16\delta_{23}\partial_{\Bar{z}_2}\delta_{12}\partial_{z_2}\partial_{\Bar{z}_2}\phi+8\delta_{12}\partial_{\Bar{z}_2}\delta_{23}\partial_{z_2}\partial_{\Bar{z}_2}\phi+4\partial_{\Bar{z}_2}\delta_{23}\partial_{z_2}\partial_{\Bar{z}_2}\delta_{12}\nonumber\\
    &-8\delta_{12}\partial_{z_2}\partial_{\Bar{z}_2}\delta_{23}\partial_{\Bar{z}_2}\phi-4\partial_{\Bar{z}_2}(\delta_{12}\partial_{z_2}\partial_{\Bar{z}_2}\delta_{23})+24\delta_{12}\delta_{23}\partial_{z_2}\partial_{\Bar{z}_2}^2\phi+4\delta_{12}\partial_{z_2}\partial_{\Bar{z}_2}^2\delta_{23}\Big]\nonumber\\
    &+\partial_{z_2}G^{z_2}_{\ zz}(z_2,\Bar{z}_2;z,\Bar{z})\Big[12\delta_{12}\partial_{\Bar{z}_2}\delta_{23}\partial_{\Bar{z}_2}\phi-6\delta_{12}\partial_{\Bar{z}_2}^2\delta_{23}\Big]+\partial_{\Bar{z}_2}G^{z_2}_{\ zz}(z_2,\Bar{z}_2;z,\Bar{z})\Big[16\delta_{12}\delta_{23}\partial_{z_2}\phi\nonumber\\
    &-4\partial_{z_2}\delta_{12}\partial_{\Bar{z}_2}\delta_{23}-4\delta_{12}\partial_{z_2}\partial_{\Bar{z}_2}\delta_{23}\Big]\Big\}+\frac{1}{16\pi^2 G}\Big\{z_2\leftrightarrow z_3\Big\}\nonumber\\
    &+\sum\limits_{\alpha=1}^{3g-3}\phi_{\alpha zz}\frac{\partial}{\partial \tau_\alpha}\braket{T_{zz}(z_1)T_{\Bar{z}\Bar{z}}(z_2)T_{\Bar{z}\Bar{z}}(z_3)}.
\end{align}

\bibliographystyle{JHEP}
\bibliography{reference.bib}

\end{document}